%% file: ms.tex
\documentclass{aa}

\usepackage{graphicx}
\usepackage{dirtytalk}
\usepackage{txfonts}
\usepackage{hyperref}
\usepackage{xcolor}
\hypersetup{
colorlinks,
linkcolor={red!80!black},
citecolor={blue!80!black},
urlcolor={blue!80!black}
}

\newenvironment{myfont}{\fontfamily{ppl}\selectfont}{\par}
\DeclareTextFontCommand{\textmyfont}{\myfont}

\newcommand{\code}[1]{\texttt{#1}}

\def\nifs{\iso{56}Ni}
\def\nife{\iso{58}Ni}

\def\cm3{cm$^{-3}$}
\def\kms{\mbox{km~s$^{-1}$}}

\def\msun{$M_{\odot}$}

\def\one{\ts {\,\sc i}}
\def\two{\ts {\,\sc ii}}
\def\three{\ts {\,\sc iii}}

\def\beq{\begin{equation}}
\def\eeq{\end{equation}}

\def\lesssim{\mathrel{\hbox{\rlap{\hbox{\lower4pt\hbox{$\sim$}}}\hbox{$<$}}}}
\def\gtrsim{\mathrel{\hbox{\rlap{\hbox{\lower4pt\hbox{$\sim$}}}\hbox{$>$}}}}

\def\one{{\,\sc i}}
\def\two{{\,\sc ii}}
\def\three{{\,\sc iii}}

\def\v1d{{\code{V1D}}}

\def\kepler{{\code{KEPLER}}}

\def\cmfgen{{\code{CMFGEN}}}

\def\photb{{\code{P-HOTB}}}

\def\nkiiopt{[Ni\two]\,$\lambda$\,7378}
\def\niidoub{[N\two]\,$\lambda\lambda$\,$6548,\,6583$}
\def\niiauroral{[N\two]\,$\lambda$\,$5755$}

\def\caiidoub{[Ca\two]\,$\lambda\lambda$\,$7291,\,7323$}
\def\caiitrip{Ca\two\,$\lambda\lambda\,8498-8662$}

\def\oidoub{[O\one]\,$\lambda\lambda$\,$6300,\,6364$}

\def\oitrip{O\one\,$\lambda\lambda$\,$7771-7775$}

\def\mgi{Mg\one]\,$\lambda\,4571$}
\def\nad{Na\one\,$\lambda\lambda\,5896,5890$}
\def\ki{K\one\,$\lambda\lambda\,7665,7699$}

\newcommand{\iso}[2]{\ensuremath{^{#1}\rm{#2}}}

\begin{document}

   \title{Modeling of the nebular-phase spectral evolution of stripped-envelope supernovae. New grids from 100 to 450 days.}

   \titlerunning{Nebular-phase evolution of SESNe}

\author{
Luc Dessart\inst{\ref{inst1}}
 \and
  D. John Hillier\inst{\ref{inst2}}
 \and
 S.E. Woosley\inst{\ref{inst3}}
 \and
Hanindyo Kuncarayakti\inst{\ref{inst4},\ref{inst5}}
 }

\institute{
    Institut d'Astrophysique de Paris, CNRS-Sorbonne Universit\'e, 98 bis boulevard Arago, F-75014 Paris, France.\label{inst1}
\and
Department of Physics and Astronomy \& Pittsburgh Particle Physics, Astrophysics, and Cosmology Center (PITT PACC),  \hfill \\
 University of Pittsburgh, 3941 O'Hara Street, Pittsburgh, PA 15260, USA.\label{inst2}
\and
Department of Astronomy and Astrophysics, University of California, Santa Cruz, CA 95064, USA\label{inst3}
\and
  Tuorla Observatory, Department of Physics and Astronomy, FI-20014 University of Turku, Finland.\label{inst4}
\and
 Finnish Centre for Astronomy with ESO (FINCA), FI-20014 University of Turku, Finland.\label{inst5}
   }

   \date{}

  \abstract{We present an extended grid of multi-epoch 1D nonlocal thermodynamic equilibrium radiative transfer calculations for nebular-phase Type Ibc supernovae (SNe) from He-star explosions. Compared to our similar work that focused on a post-explosion epoch of 200\,d, we study the spectral evolution from 100 to about 450\,d and augment the model set with progenitors that were evolved without wind mass loss. Models with the same final, preSN mass have similar yields and produce essentially the same emergent spectra. Hence, the uncertain progenitor mass loss history compromises the inference of the initial, main sequence mass. This shortcoming does not affect Type IIb SNe in which mass-loss has left a small residual H-rich envelope in the progenitor star at core collapse, and hence an intact He core. However, our 1D models with a different preSN mass tend to yield widely different spectra,  as seen through variations in the strong emission lines due to \niidoub, \oidoub, \caiidoub, \nkiiopt, and the forest of Fe\two\ lines below 5500\,\AA. At the lower mass end, the ejecta are He rich and at 100\,d  cool through He\one, N\two, Ca\two, and Fe\two\ lines, with N\two\ and Fe\two\ dominating at 450\,d. These models, associated with He giants, conflict with observed SNe Ib, which typically lack strong N\two\ emission.  Instead they may lead to SNe Ibn or, because of additional stripping by a companion star, ultra-stripped SNe Ic. In contrast, for higher preSN masses, the ejecta are progressively He poor and cool at 100\,d through O\one, Ca\two, and Fe\two\ lines, with O\one\ and Ca\two\ dominating at 450\,d. Nonuniform, aspherical, large-scale mixing rather than composition differences likely determines the SN type at intermediate preSN masses. Variations in clumping, mixing, as well as departures from spherical symmetry would increase the spectral diversity but also introduce additional degeneracies. More robust predictions from spectral modeling thus require a careful attention to the initial conditions, by incorporating the salient features of physically-consistent 3D explosion models.
}


   \maketitle


\section{Introduction}

In a previous work \citep[hereafter D21]{dessart_snibc_21}, we performed non-local thermodynamic equilibrium (nonLTE) radiative transfer calculations with the code \cmfgen\ \citep{HD12} employing the He-star progenitor and explosion models of \citet{woosley_he_19} and \citet{ertl_ibc_20}. These explorations were focused on a SN age of 200\,d, and complemented earlier calculations for similar He-star models that focused on the photospheric phase \citep{dessart_snibc_20,woosley_ibc_21}. Here we extend the study of D21 to cover the nebular phase from 100 to about 450\,d after explosion for the He-star models of D21 with initial mass between 2.6 and 12.0\,\msun\ (corresponding to masses between 13.85 and 35.74\,\msun\ on the H zero age main sequence).

The aim of this work is to compute Type Ib and Ic SN observables associated with He star models covering from the lowest mass massive stars undergoing core collapse up to higher mass He stars just below the threshold for black hole formation and implosion \citep{ertl_ibc_20}. Objects at the lower mass end are favored by the initial mass function but their complicated physics (nuclear flashes) and special properties (large radii) might lead to peculiar SNe rather than standard Type Ib SNe (see, for example, the recent works of  \citealt{yoon_ibc_15}; \citealt{kleiser_kasen_hegiant_18}; or \citealt{woosley_he_19}). At the high mass end, He stars die without a He-rich shell (He/N or He/C) and represent ideal candidates for Type Ic SNe \citep{dessart_snibc_20}. Because of the fundamental asymmetric nature of the neutrino-driven explosion mechanism and the resulting 3D asymmetric, distribution of elements \citep{wongwathanarat_15_3d,gabler_3dsn_21}, non-thermal excitation of He may vary from SN to SN and with viewing angle. Thus objects with similar preSN mass of say 6\,\msun\ may appear as Type Ib or Type Ic \citep{d12_snibc}.   In the present work, spherical symmetry is assumed so that the division between Type Ib and Type Ic is sharp (probably too much so), and is tied to the presence or absence of the He/N shell in the progenitor at explosion \citep{dessart_snibc_20}.

Few works have been devoted to the numerical modeling of stripped-envelope SNe at nebular epochs. NonLTE models have been produced for Type IIb SNe  \citep{maurer_iib_neb_10,jerkstrand_15_iib,ergon_11dh_22}, Type Ic SNe \citep{mazzali_etal_07gr}, GRB/SNe like 1998bw  \citep{mazzali_98bw_01,dessart_98bw_17}, or superluminous Type Ic SNe \citep{jerkstrand_slsnic_17,D19}. But no systematic grid has ever been explored in detail for Type Ib and Ic SNe apart from D21. Indeed, these earlier simulations have often been one-zone models with no global and consistent connection to a progenitor and an explosion model, or they were based on single-star progenitor models whilst a binary progenitor channel was invoked. Binarity is probably essential for the production of Type Ib and Ic SNe (see, e.g., \citealt{wheeler_levreault_85}; \citealt{ensman_woosley_88}; \citealt{podsiadlowski_92}) so employing binary-star progenitor models, together with their explosive result, appears mandatory in this context (see, e.g., \citealt{laplace_evol_21}). Here, we document the nebular-phase properties of a grid of He-star models covering the full range of masses for Type Ib and Ic SNe.  These He-star models are thought to arise from binary mass transfer shortly after central helium ignition and thus implicitly require a binary progenitor (the subsequent evolution ignores any binary effect). Furthermore, we assume a solar metallicity and ignore rotation, so further work is needed to investigate the impact of these aspects on Type Ib and Ic SN observables.

In the next section, we briefly present the numerical setup for the present study. All details about the progenitor and explosion models as well as the radiative transfer approach are to be found in D21. In Section~\ref{sect_degen}, we compare our results for two models of different initial mass but the same preSN mass and thus highlight some of the degeneracies of nebular-phase spectra. The photometric evolution for our grid of models is described in Section~\ref{sect_phot}. We then discuss in detail and in turn our results for the main spectral landmarks of Type Ibc nebular spectra, namely the emission lines \niidoub, \oidoub, \caiidoub, \nkiiopt, and the forest of Fe\two\ lines below about 5500\,\AA. In Section~\ref{sect_conc}, we discuss the implications of these results and present our conclusions. Comparison to observations are deferred to Kuncarayakti et al. (in preparation). All model spectra are publicly available online.\footnote{\url{https://zenodo.org/communities/snrt}}

\section{Numerical setup}
\label{sect_prep}

In this work, we continue and extend the radiative transfer modeling of He-star explosions presented in D21. Two sets of progenitor models are used. The first set corresponds to the He-star models of D21 with initial masses between 2.6 and 12.0\,\msun\ and evolved with a nominal mass loss rate \citep{yoon_wr_17}. These were given the names he2p6, he2p9, ...,  he12p0. This set is augmented with He-star models evolved without mass loss in order to consider progenitors in which the mass transfer phase leading to the H-rich envelope stripping occurs just before core collapse. Such models are suitable for progenitors leading to Type IIb SNe endowed with trace amounts of H-rich material.\footnote{At nebular times, this H-rich material is invisible (\citealt{jerkstrand_15_iib}, \citealt{ergon_11dh_22}, D21), unless the ejecta interact with circumstellar material \citep{matheson_93j_00a, d13_late_sn2p}.}  To ensure moderate ejecta masses compatible with observations (for example, see \citealt{ensman_woosley_88} or \citealt{drout_11_ibc}) we selected only a few masses  in the range 2.5 to 4.5\,\msun, spaced every 0.5\,\msun. These models, evolved and exploded in the same fashion as the other set, are named he2p5MdZ etc -- the suffix ``MdZ'' stands for MdotZero. This second set of models is used to assess the degeneracy of nebular phase spectra when determining yields or inferring the progenitor mass. As demonstrated below and already envisioned, the spectral properties are primarily dependent on the preSN mass, which depends on both the initial star mass and the uncertain wind mass loss. A summary of the ejecta properties for both model sets is provided in Table~\ref{tab_prog}.

The second change relative to D21 is that all models are evolved from 100 to about 450\,d while only models he5p0, he6p0, and he8p0 were evolved (and only from 200 to 450\,d) in D21. We thus provide a more extended nebular-phase evolution and for a wider range of He-star explosion models than currently available in the literature -- this new set of simulations represents the largest grid of radiative transfer models for SNe Ibc at nebular times. Ideally, one would have evolved the models from a much earlier time, say 1\,d after explosion, but the present simulations, which employ the shuffled-shell technique for macroscopic mixing \citep{DH20_shuffle}, require high resolution (with typically 350 radial points) and are thus extremely costly.\footnote{The shuffled ejecta corresponding to models he2p6 to he12p0 used in this study are identical to those used in D21 -- they are the same ejecta, only simulated over a different time span. The same procedure was applied to the unmixed ejecta models from progenitors evolved without mass loss. The approach in all cases was to split all shells (except the outermost shell) into three subshells and distribute them in the same order, starting from the innermost ejecta layer. We thus obtain a pattern of alternating shells of different composition repeated three times, connecting in the outer ejecta with the outermost shell that we excluded from the shuffling -- see D21 for details and illustrations.} Furthermore, distinguishing between microscopic and macroscopic mixing during the photospheric phase is less critical because the plasma behaves similarly to a blackbody (i.e., a thermal emitter moderately influenced by line opacity and emissivity rather than a nebula cooled by lines). Chemical inhomogeneities and clumping nonetheless affect the SN radiation and gas properties (e.g., through a change in the radiative diffusion time) during the photospheric phase \citep{d18_fcl, dessart_audit_rhd_3d_19,ergon_11dh_22}.

\input{model_composition_table.tex}

D21 presented an in-depth study of the radiative transfer at nebular times in Type Ib and Ic SN ejecta, but with a primary focus on a SN age of 200\,d. Here, we compute the evolution at 20 epochs for 15 models, which corresponds to a total of 300 models. To avoid a lengthy presentation and repetition of results already presented in D21, the paper focuses on the evolution of key line diagnostics associated with the cooling of the four main ejecta shells (i.e., He, O, Si, or Fe rich) with only a brief description of the evolution of the optical spectra for a few representative models. The complete spectral evolution for the full set of models is presented in the appendix.

\begin{figure*}
\centering
\includegraphics[width=0.45\hsize]{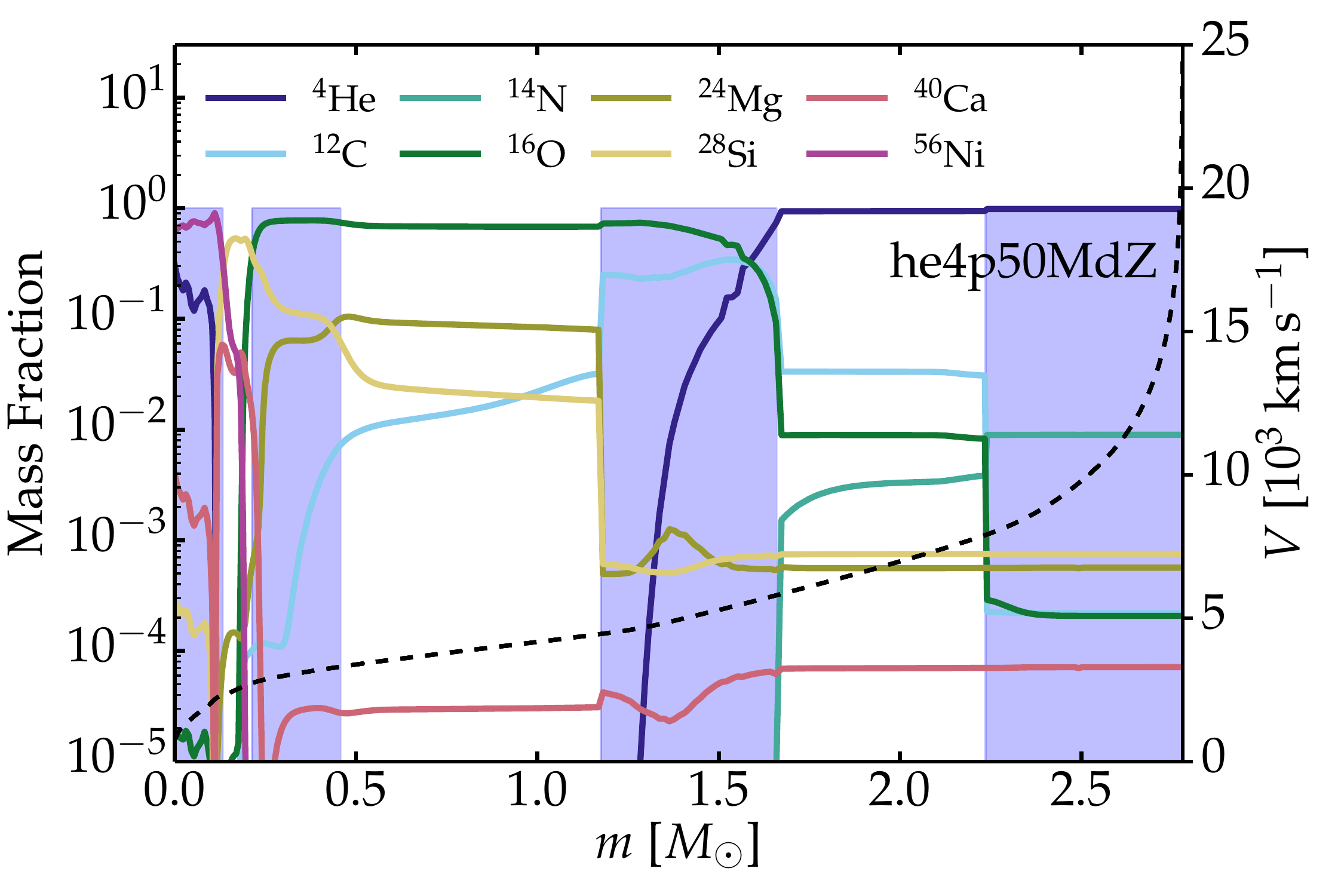}
\includegraphics[width=0.45\hsize]{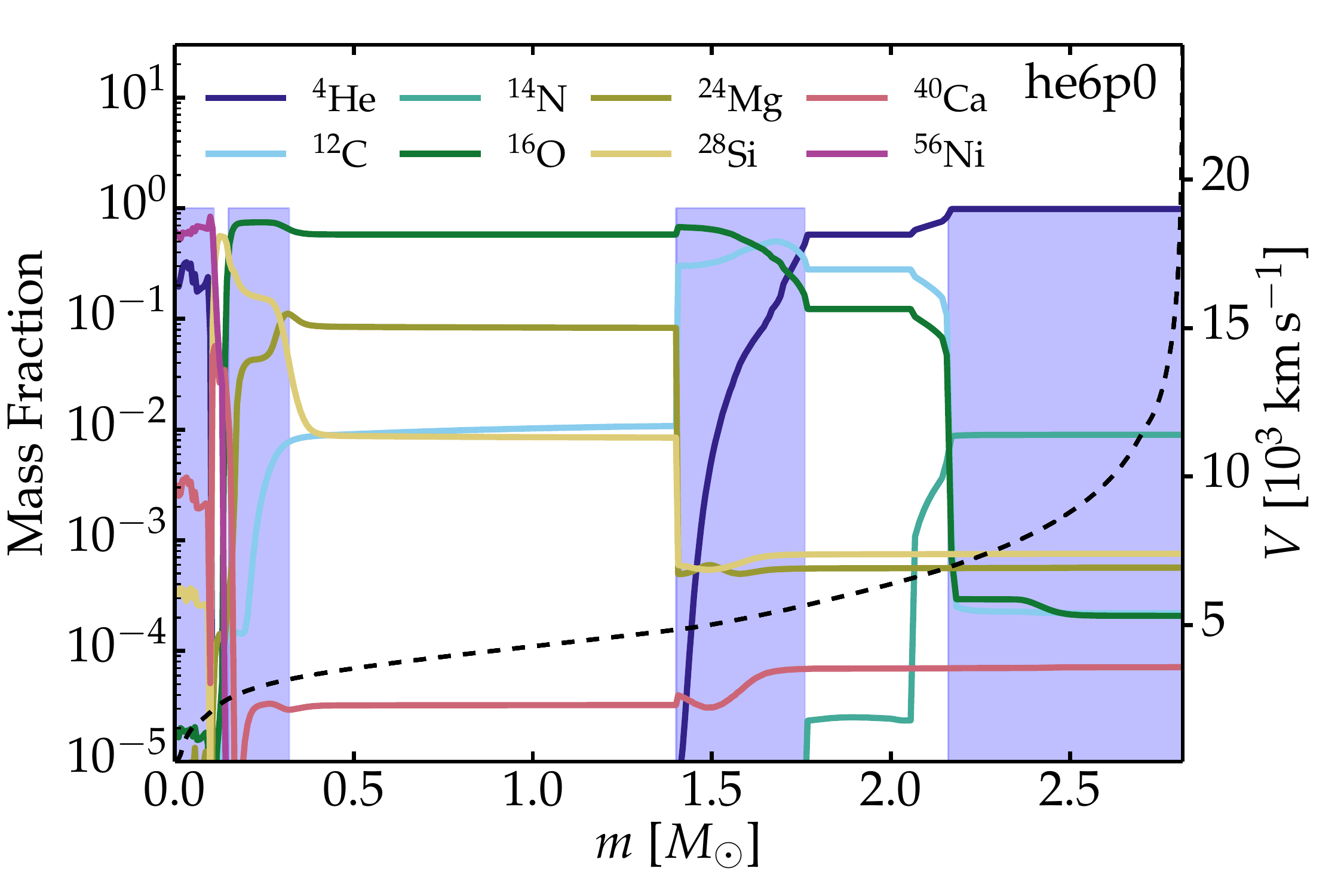}
\caption{Comparison of the he4p5MdZ (left; $M_{\rm preSN}=$\,4.5\,\msun) and he6p0 (right; $M_{\rm preSN}=$\,4.44\,\msun) unmixed ejecta models. The alternating shades of blue and white indicate the successive shells in the original unmixed ejecta. Starting from the innermost ejecta layer, one progresses outwards through the Fe/He, Si/S, O/Si, O/Ne/Mg, O/C, He/C, and He/N shells.
\label{fig_ej_init}
}
\end{figure*}

\begin{figure}[ht!]
\centering
\includegraphics[width=\hsize]{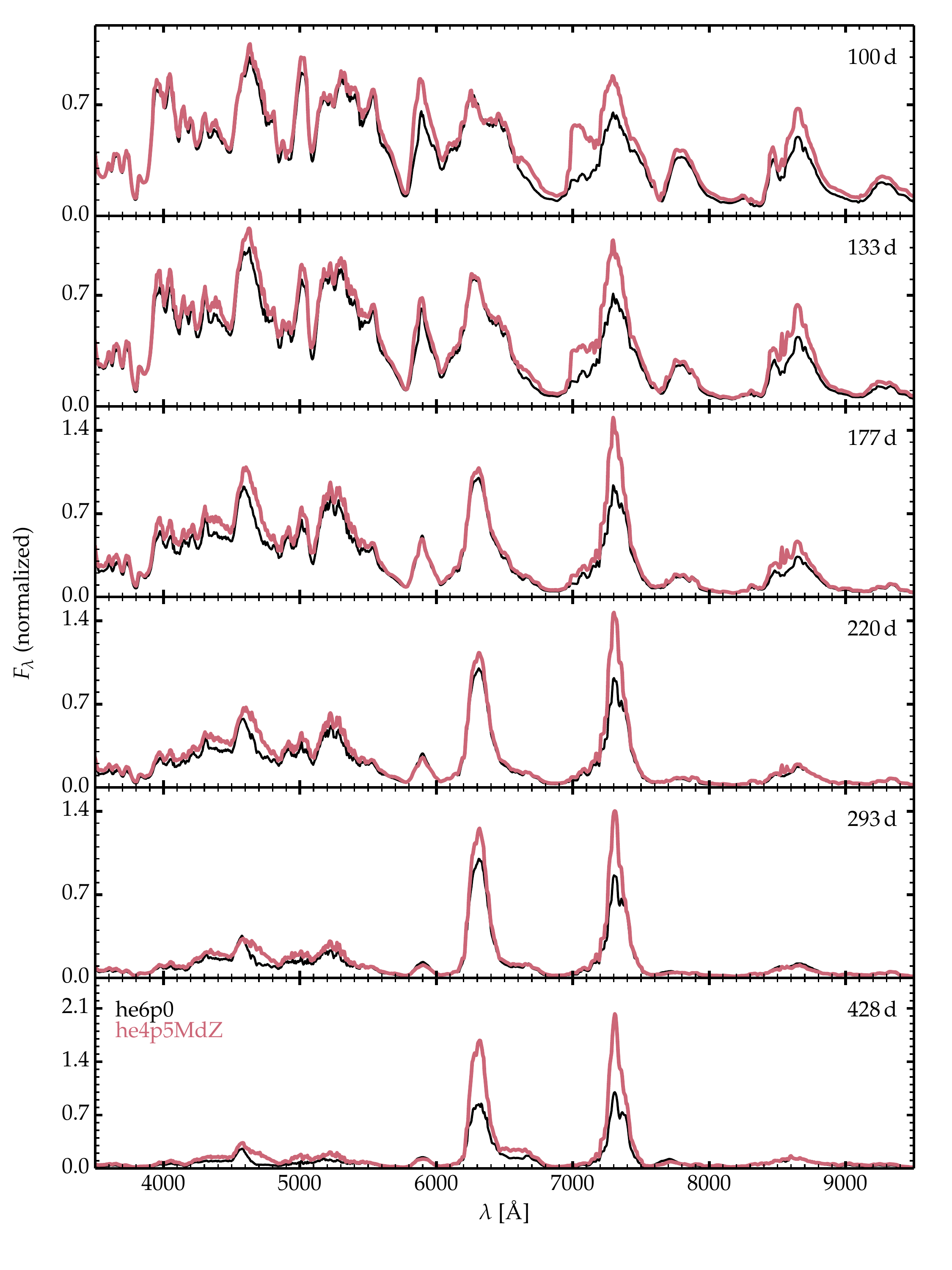}
\caption{Comparison of multi-epoch optical spectra for models he6p0 (black; with nominal mass loss) and he4p5MdZ (red; progenitor evolved with no mass loss) from 100 to 428\,d after explosion. At each epoch, the same normalization is applied to both spectra.
\label{fig_he4p5MdZ_he6}
}
\end{figure}

\section{Properties of nebular phase spectra reflect primarily the preSN mass}
\label{sect_degen}

The main stellar parameter sought for in nebular-phase spectroscopic modeling is the mass of the progenitor star on the main sequence. Oxygen is often used in this context as a diagnostic since its abundance is a rising function of He core mass in single star models \citep{whw02}, and with final (i.e., preSN) mass in He-star models \citep{woosley_he_19}. This property has been extensively used to constrain the progenitor masses of Type II SNe (see, for example, \citealt{maguire_2p_12,jerkstrand_04et_12,silverman_neb_17,wynn_20tlf_22}). The method has also been applied to stripped-envelope SNe but, as we demonstrate here, the method is flawed in SNe Ib and Ic because of preSN mass loss (SNe IIb are not affected because mass-loss has left a small residual H-rich envelope, and hence an intact He core in the progenitor at core collapse).

D21 showed the close correspondence between the O mass and the preSN mass (for H-deficient stars) or the He-core mass (for H-rich stars) for a wide range of binary-star and single-star models. Using the O yield to infer the original, main sequence mass of H-deficient stars is, however, difficult because of the influence of mass loss on the bare He core once the H-rich envelope is gone. This  uncertain impact of mass loss compromises any robust determination of the initial mass of Type Ib and Ic progenitors -- one may only set a lower bound on the initial mass and infer their preSN mass from nebular-phase observations.

This issue with mass loss does not affect Type II and IIb SNe since their He core remains enshrouded throughout the progenitor life. In addition to these considerations, the O yield may be further affected by the processes of convection, semi-convection, overshoot, or rotation, for which we lack a robust theory (e.g., convection) or good observational constraints (e.g., rotation).

As will be demonstrated by the following discussion, our two sets of models with and without mass loss teach us a number of important lessons. First, the models confirm the good correlation between O-yield and preSN mass (Table~\ref{tab_prog}, and particularly Table~\ref{tab_o_yields}; see also Section~3 and Fig.~3 in D21). Second, the composition structure of models with the same preSN mass exhibits subtle differences  -- different shells have different mass and exhibit different compositions. Third, different progenitor masses on the He zero-age main sequence can lead to similar nebular spectra. Fourth, while the \oidoub\ line strength may provide a robust indication of the O yield, its connection to the initial mass of the progenitor on the H or He zero age main sequence is more tenuous.

\begin{table}[h]
\caption{Comparison of oxygen yields \label{tab_o_yields}}
\begin{center}
\begin{tabular}{lcc}
\hline
Model  &  $M_{\rm preSN}$  &    O yield \\
            &  [\msun]        &    [\msun] \\
\hline
he2p9 & 2.37 & 0.05  \\
he2p5dZ & 2.50 & 0.092 \\
he3p3 & 2.67 &  0.15 \\
\\
he4p0     &     3.16  & 0.31 \\
he3p0MdZ  & 3.00  &  0.21 \\
\\
he3p5MdZ  & 3.50  &  0.43\\
he4p5       &  3.49  & 0.42 \\
he5p0       &  3.81  &  0.59 \\
\\
he4p5MdZ & 4.50  & 0.96 \\
he6p0        &  4.44 &  0.97 \\
\hline
\end{tabular}
\end{center}
\end{table}

Figure~\ref{fig_ej_init} shows the unmixed composition (i.e., prior to the shell shuffling used to enforce some macroscopic mixing) for models he4p5MdZ and he6p0. While the preSN mass and O-yield are identical within a few percent, model he4p5MdZ has a bigger O/Si shell, a smaller O/Ne/Mg shell, a  bigger O/C shell, and its He/C shell has a lower C and O mass fraction than in the model he6p0. Inferring such subtle differences from nebular phase spectroscopy is probably impossible, in part because macroscopic mixing transforms this clean shell structure into a complicated 3D structure. The mass range occupied by the O-rich shell ($\sim$\,0.2 to $\sim$\,1.7\,\msun\ above the innermost ejecta layer at 1.6--1.7\,\msun) and the He-rich shell (the layers above the O-rich shell) is nonetheless comparable in both models. Finally, the core structure differs. Although the mass cut in model he4p5MdZ was located further out (1.71 instead of 1.60\,\msun\ in model he6p0), its compactness and its explosion energy are greater ($E_{\rm kin}=$\,1.2\,foe instead of 1.1\,foe), causing a more extensive explosive nucleosynthesis (0.1 versus 0.07\,\msun\ -- a greater \nifs\ mass is also found in other MdZ models relative to He star models evolved with mass loss and having the same preSN mass; see Table~\ref{tab_prog}).

The differences in composition structure can be explained. Despite the similar final mass, the models he6p0 and he4p5MdZ are different stars with different evolutionary histories. In the absence of wind mass loss (model he4p5MdZ), the mass stays fixed during helium burning and the convective core extends out to some point appropriate to that mass and stays there. However, when wind mass loss operates, the star mass shrinks and the helium convective core recedes accordingly. The  resulting entropy profile and location of burning shells, in particular the helium burning shell, are altered. The relatively modest differences in final composition profiles suggest that the mass lost was not a large fraction of the initial mass. Furthermore, the difference in explosion properties reflects the differences in their preSN core structure. For example, the radius encompassing 2.5\,\msun\ in he4p5MdZ is $1.56 \times 10^9$\,cm. For he6p0 it is $2.58 \times 10^9$\,cm. This results in different efficiencies of the explosion engine.

Figure~\ref{fig_he4p5MdZ_he6} compares the \cmfgen\ spectra of model he4p5MdZ and he6p0 at multiple epochs spanning the range 100 to 428\,d. Despite the different progenitor masses on the He zero-age main sequence, the similarity in ejecta composition, mass, and kinetic energy lead to very similar spectra at all epochs. Some differences are however noticeable at certain epochs and locations. More power is absorbed in the more He-pure He-rich shell of model he4p5MdZ (Fig.~\ref{fig_ej_init}), which leads to stronger He\one\ lines at 100 and 133\,d -- no obvious He\one\ line is present at later times in the optical range. Furthermore, the greater \nifs\ mass in model he4p5MdZ (0.1 versus 0.07\,\msun) leads to a stronger \caiidoub\ since this is the main coolant of the Fe/He and Si/S shells (D21; both shells are bigger in model he4p5MdZ; see left column of Fig.~\ref{fig_ej_init}). This emphasizes that the ratio of emission fluxes associated with \oidoub\ and \caiidoub\ is not a straightforward progenitor mass diagnostic -- the O yield scales with the preSN mass and preSN nucleosynthesis while the \caiidoub\ line strength scales with the amount of explosively nucleosynthesized material, which depends on numerous progenitor characteristics other than preSN mass (i.e., compactness, explosion energy, or mass cut).

At the latest epoch, model he4p5MdZ shows much stronger \oidoub\ and \caiidoub\ line emission than model he6p0. The relative fraction of power absorbed in the various shells is very similar in both models but because of its greater \nifs\ mass, all powers are scaled up by $\sim$\,50\% in model he4p5MdZ so that line fluxes are also $\sim$\,50\% stronger in model he4p5MdZ. Renormalizing the spectra so that both models have the same decay-power absorbed reduces the offset between the model spectra but some offset remains. This can be explained by differences in composition and ionization. For example, model he4p5MdZ has a higher $E_{\rm kin}/M_{\rm ej}$  which leads to a slightly higher ionization in the O-rich shell -- the more recombined O-rich material in the he6p0 model cools in part through \nad, causing a reduction in the flux of \oidoub. This shows the complexity of inferring ejecta properties and composition from nebular-phase spectra, even in a controlled experiment based on two very similar ejecta models.

To conclude, this comparison emphasizes that while the \oidoub\ line strength may provide a robust indication of the O yield, the connection to the initial mass (whether that is the H or the He zero age main sequence) cannot be established precisely. Only a range of progenitor masses may be inferred based on various assumptions concerning the preSN mass loss history. It also shows that one may produce a variety of progenitor models for a given nebular-phase spectrum (the present comparison also ignores any departure from spherical symmetry or the effect of clumping).

\begin{figure}[ht!]
\centering
\includegraphics[width=\hsize]{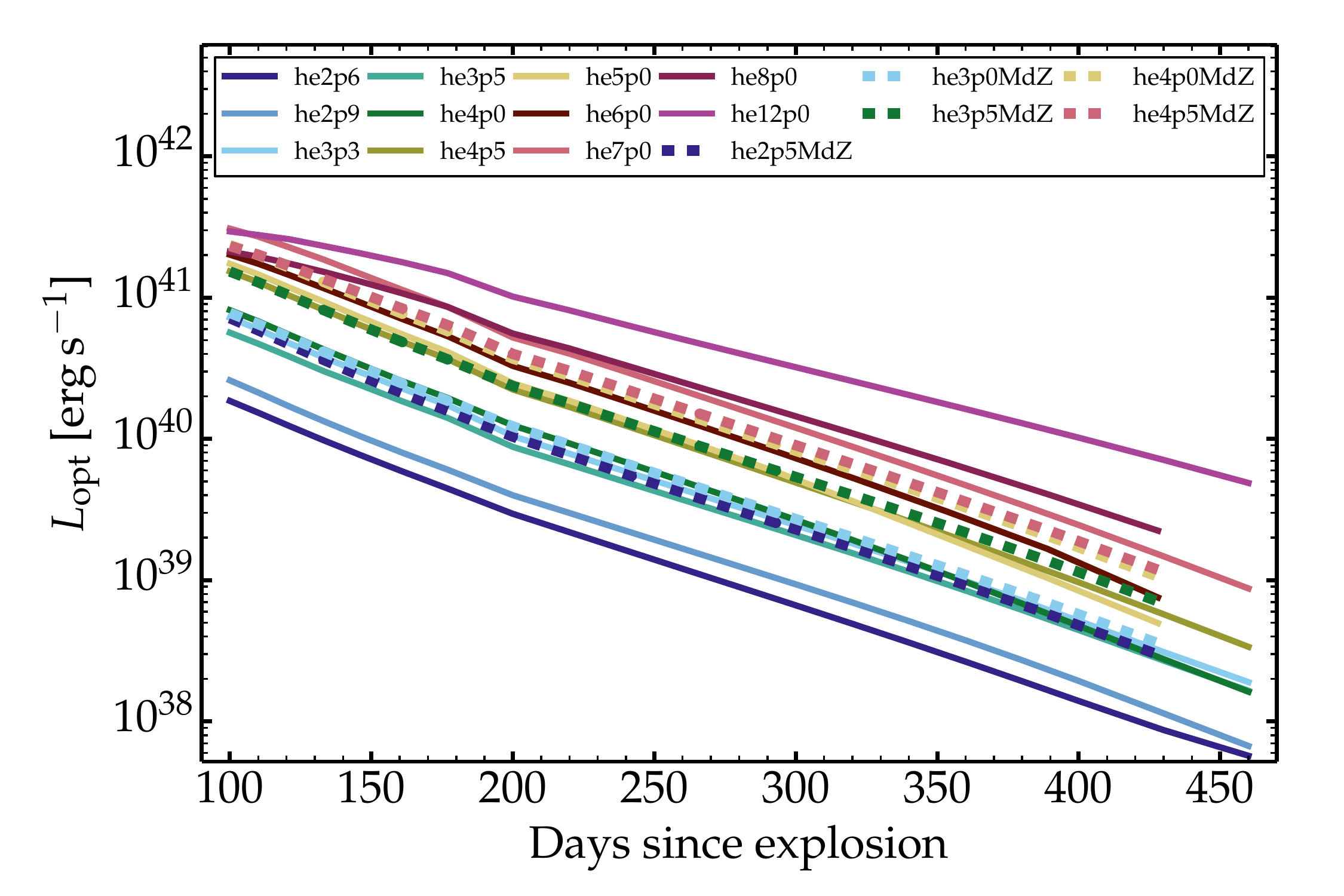}
\caption{Evolution of the optical luminosity for our set of He-star explosion models evolved with and without mass loss. At 100 days the models vary in flux by over an order of magnitude, while at 450 days the variation in optical flux is nearly two orders of magnitude.
\label{fig_lopt}
}
\end{figure}

\section{Photometric evolution}
\label{sect_phot}

Figure~\ref{fig_lopt} shows the evolution of the optical luminosity (corresponding to wavelengths between 3500 and 9500\,\AA) from 100 to about 450\,d after explosion for the full set of models. The evolution of the bolometric luminosity and the fraction of that power falling in the UV (wavelengths below 3500\,\AA), optical, and IR (wavelengths beyond 9500\,\AA) ranges is shown in Fig.~\ref{fig_lbol_frac}.

As for SNe II at nebular epochs \citep{D21_sn2p_neb}, these He-star explosion models radiate about 70\% of their luminosity in the optical range (it is about 10\% in the UV, and about 20\% in the IR; Fig.~\ref{fig_lbol_frac}). Relative to Type II SN models of different masses \citep{D21_sn2p_neb}, there is here a much greater scatter. Lower mass progenitors, because of their greater He content, are systematically bluer, with more flux emerging in the UV (up to 15\% for model he2p6) and less in the optical and IR. Higher mass models (especially he8p0 and he12p0) are cooler and more recombined and exhibit their strongest lines in the optical where they radiate up to 80\% of their total flux with just a few percent falling in the UV.

This relative constant distribution of the SN luminosity between UV, optical, and IR ranges occurs despite the drop in bolometric luminosity by a factor of 50 to 100 between 100 and 450\,d.  As already discussed in D21, $\gamma$-ray leakage progressively increases and modulates the relative contribution from positrons, which we assume are absorbed locally (Fig.~\ref{fig_edep_evol}). In model he12p0, there is nearly full $\gamma$-ray trapping at 100\,d but only 30\% trapping at 450\,d. At the opposite end of the mass range, model he2p9 traps 30\% of emitted $\gamma$ rays at 100\,d. For all light models, the fractional decay power absorbed at 450\,d is 6--10\%, with therefore as much as 50\% of the total decay power coming from positrons. In such models, this implies that 50\% of the emission originates from the ejecta regions rich in \nifs\ initially.

\begin{figure*}[h!]
\centering
\includegraphics[width=0.42\hsize]{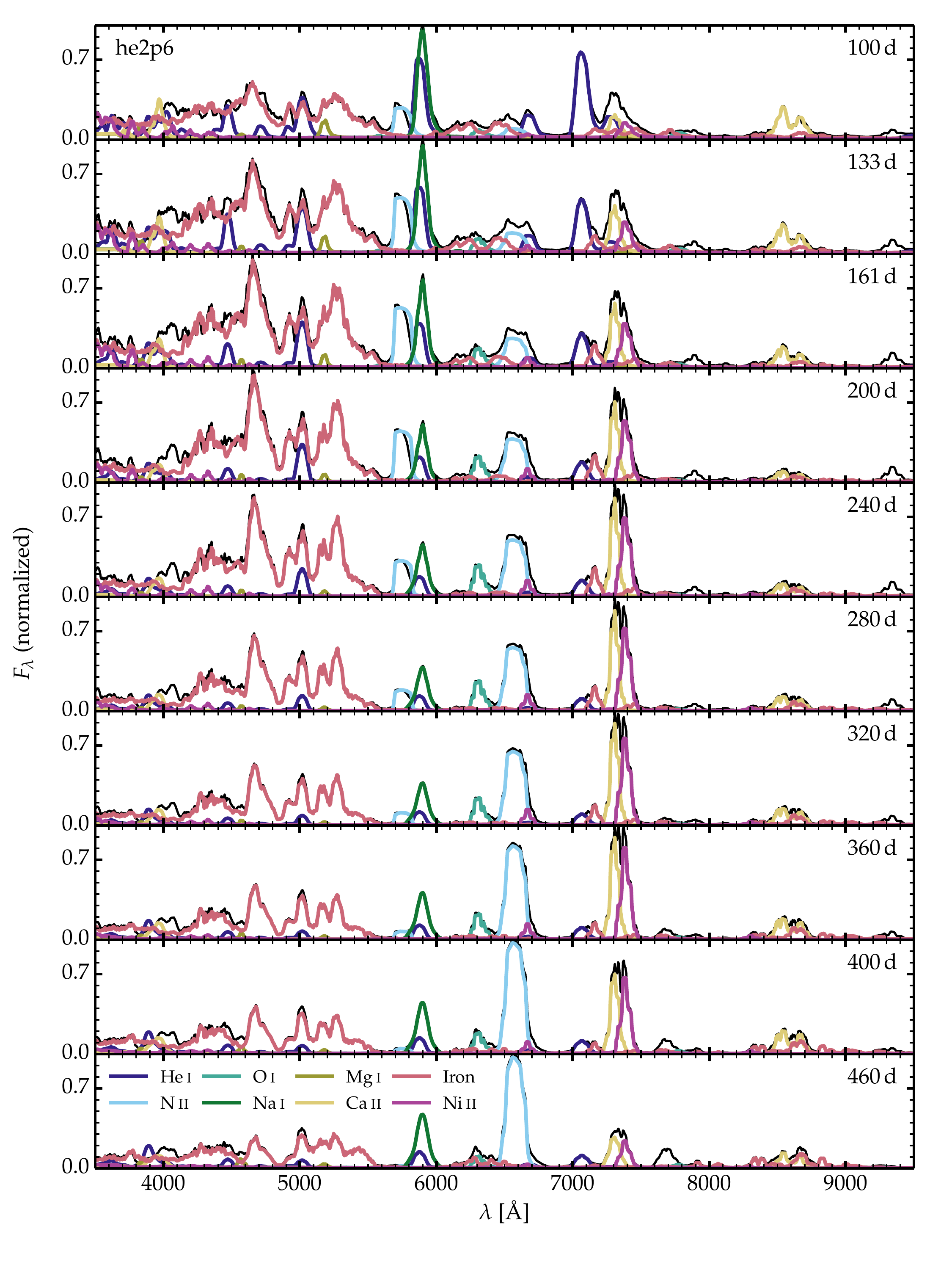}
\includegraphics[width=0.42\hsize]{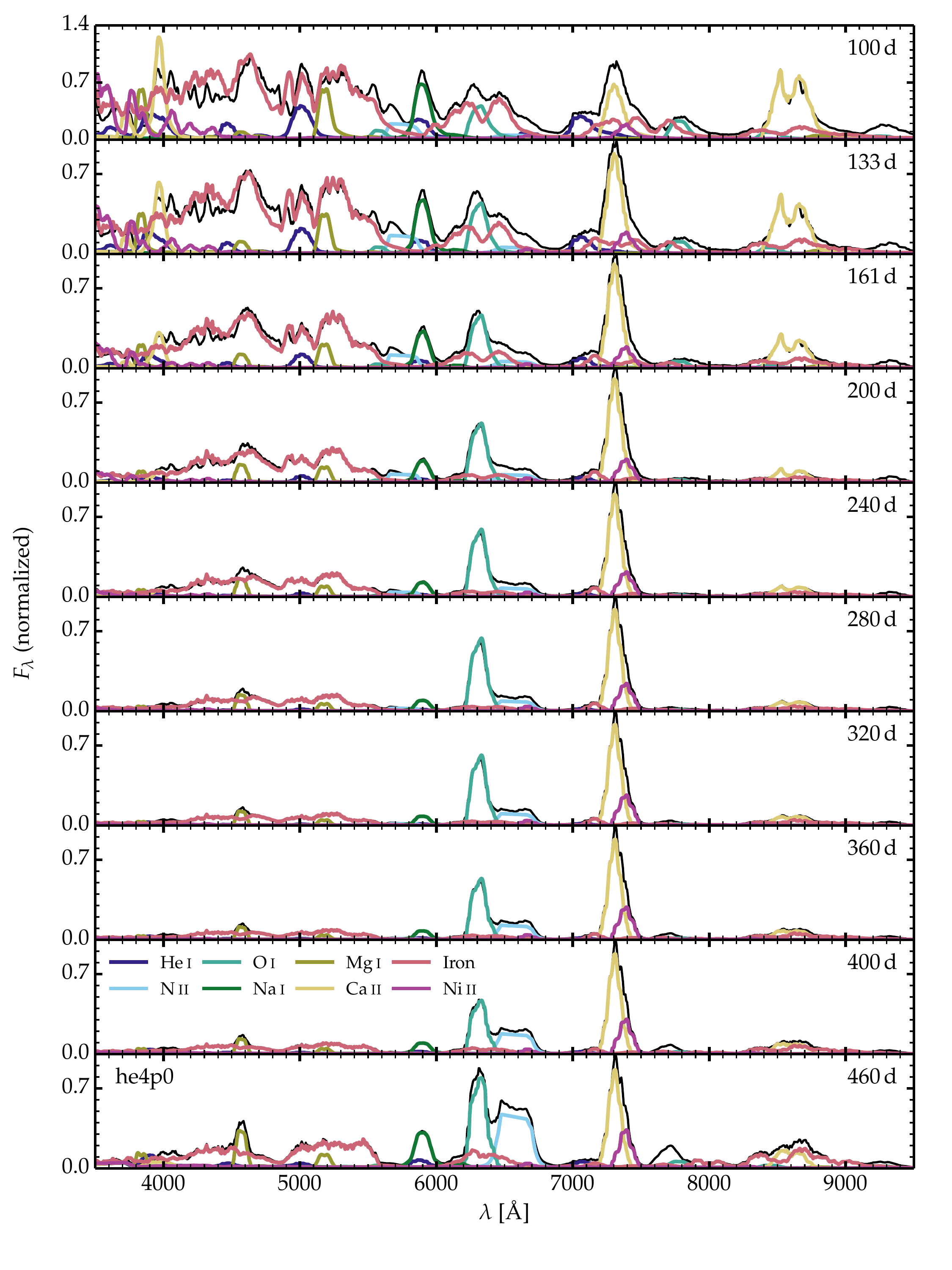}
\includegraphics[width=0.42\hsize]{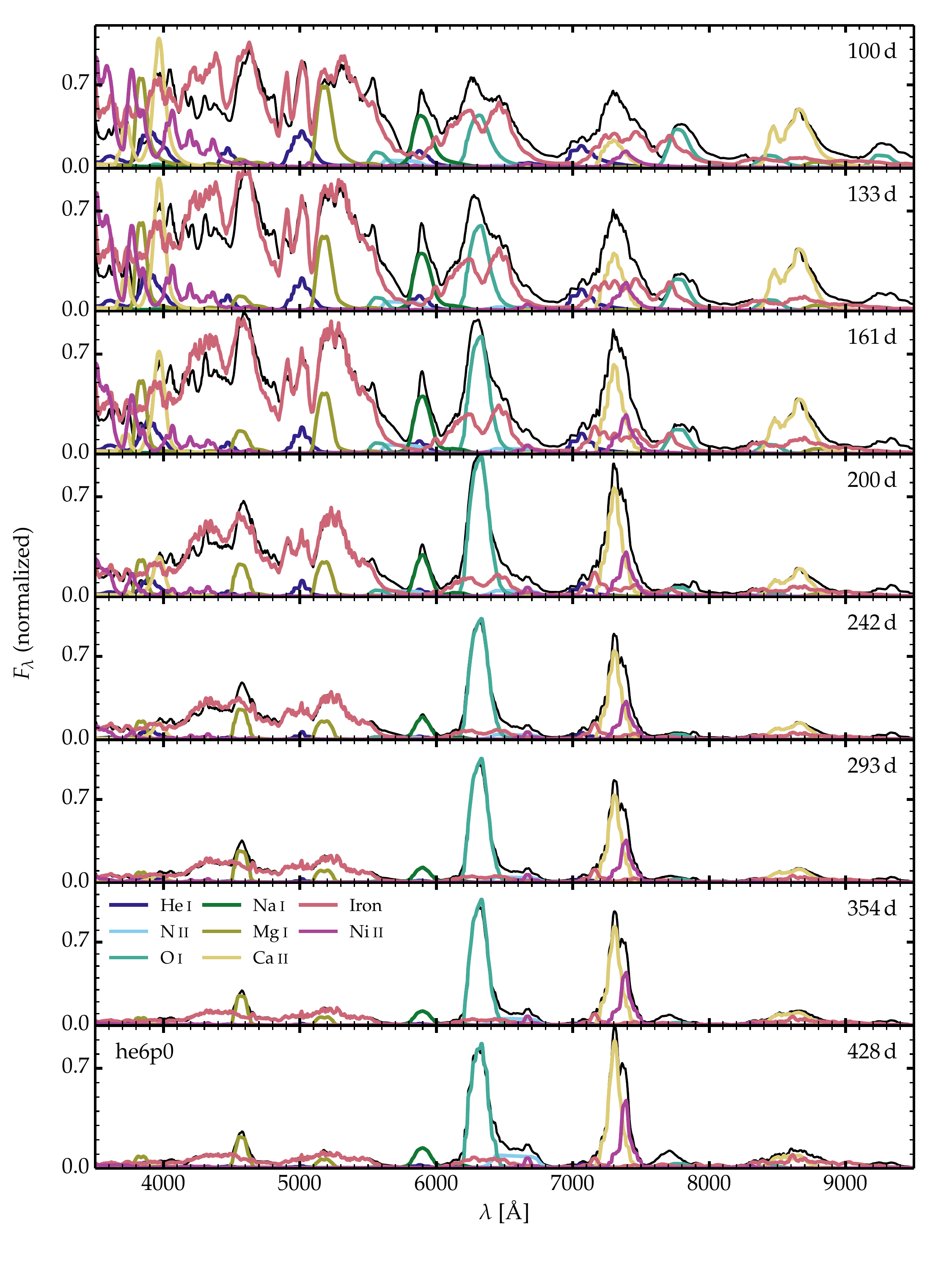}
\includegraphics[width=0.42\hsize]{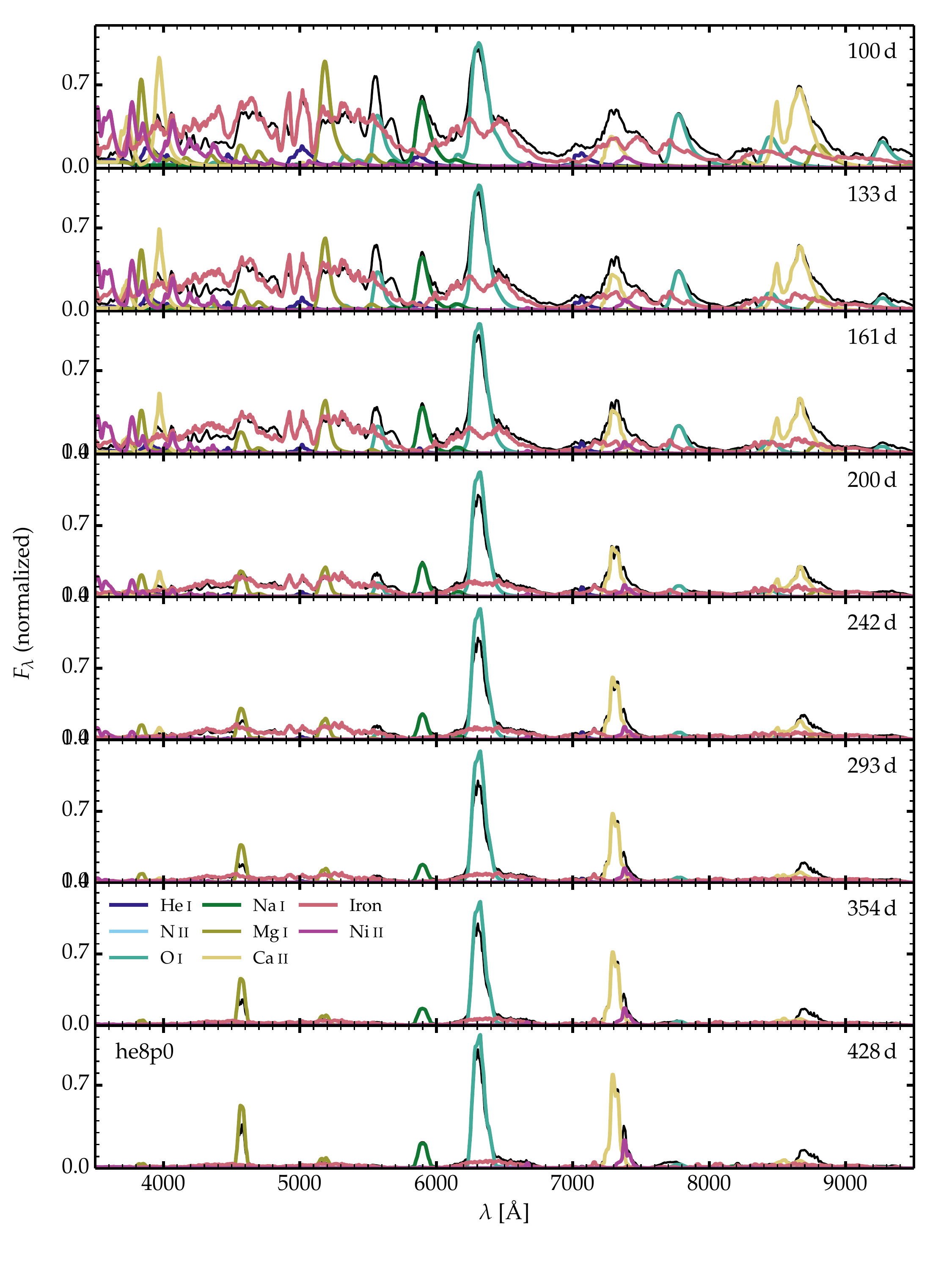}
\caption{Evolution of the optical spectrum of models he2p6 (top left), he4p0 (top right), he6p0 (bottom left), and he8p0 (bottom right) from 100 to about 450\,d (black). Also plotted is the contribution from bound-bound transitions associated with various ions (``Iron'' corresponds to a simulation of the spectrum that includes all Fe\one, Fe\two, and Fe\three\ bound-bound transitions) --- these individual contributions dominate in turn in distinct spectral regions so that the black line giving the total flux, which is exclusively an emission line spectrum, is hardly visible at any epoch. \niidoub\ dominates the emission near 6500\,\AA\ in he2p6, is present in he4p0 at later epochs, but absent in the higher preSN mass models. In the later two models, \oidoub\ is the prominent feature near 6300\,\AA.
\label{fig_evol}
}
\end{figure*}

\section{Spectral evolution for four He-star progenitor masses}
\label{sect_res}

The He-star explosion models discussed in D21 reveal a wide range of nebular-phase spectral properties. At 200\,d after explosion, we identified three different families with prototypes the he3p3, the he5p0, and the he8p0 models. These distinct models also exhibit different evolutions from 100 to 450\,d. To avoid duplication with D21 and to cover more finely the differences between models, we show in Fig.~\ref{fig_evol} the optical spectral evolution for models he2p6, he4p0, he6p0, and he8p0. This figure also details the contributions from bound-bound transitions associated with He\one, N\two, O\one, Na\one, Mg\one, Ca\two, ``Iron'' (i.e., the synthetic spectrum computed with a formal solution includes simultaneously Fe\one, Fe\two, and Fe\three\ bound-bound transitions) and Ni\two. In each model, similar lines are seen at all epochs but their strengths varies greatly in time. In this section, we present the main spectral characteristics and defer to the following sections their interpretation (see also D21).

Model he2p6 (top left panel of Fig.~\ref{fig_evol}) is at the low-mass end of our model set and exhibits spectral properties unlike that typically observed for Type Ib and Ic SNe. At 100\,d, the spectrum exhibits strong He\one\ recombination lines at 5875, 6678, 7065\,\AA, \niiauroral\  and \niidoub, a few O\one\ lines including  weak \oidoub, \nad\ (which overlaps with He\one\,5875\,\AA), Ca\two\ lines with the \caiidoub\ doublet (which merges into a single broad feature with the adjacent emission from \nkiiopt), and finally an extended Fe emission (primarily due to Fe\two) between 4000 and 5500\,\AA. As time passes, the He\one\ lines weaken (He\one\,6678\,\AA\ soon vanishes but the lines at 5875 and 7068 persist throughout the evolution). \niiauroral\ weakens while \niidoub\ strengthens and becomes the strongest line in the optical range at $\gtrsim$\,400\,d. \nad\ eventually becomes the primary contributor to the 5900\,\AA\ emission feature. At most epochs the \caiidoub\ doublet is strong except at the earliest and latest epochs. Iron emission abates a little in the course of the evolution but remains quite strong.

Model he4p0 (top right panel of Fig.~\ref{fig_evol}) shares a number of properties with model he2p6 but the iron emission is much weaker after about 200\,d, the \niidoub\ doublet (with a broad and boxy profile) follows a similar evolution but never becomes as strong, the \oidoub\ doublet is much stronger, the \caiidoub\ is strong at all times (here with a smaller contribution from \nkiiopt) and stronger than \oidoub\ at all times. \mgi\ is now predicted at most epochs but invisible before about 400\,d, in part because of blanketing by overlapping Fe\two\ transitions. The \caiitrip\ is quite strong initially but vanishes after about 200\,d.

Model he6p0 (bottom left panel of Fig.~\ref{fig_evol}) is analogous to model he4p0 but shows a stronger \oidoub\ at all times, which now rivals in strength with the \caiidoub\ doublet, while the \niidoub\ is now much weaker and hard to distinguish. The iron emission is strong from 100 to 200\,d but progressively weakens until the last epoch. In model he6p0, \nad\ is the sole contributor to the 5900\,\AA\ feature except at the earliest epochs.

Model he8p0 (bottom right panel of Fig.~\ref{fig_evol}) shows a weaker iron emission than all other models, and this iron emission eventually vanishes at late times when the model cools primarily through \mgi, \nad, \oidoub, and \caiidoub. As in model he6p0, the \oitrip\ line is strong at early times. Apart from the earlier epochs with Fe\two\ emission, the spectrum changes little beyond 200\,d. Of these four models, he8p0 shows the strongest \oidoub, which is also the strongest emission in the optical at all times between 100 and 428\,d.
We also note that in all models after about 400\,d, the resonance transition \ki\ strengthens -- this should not be mistaken with a blueshifted emission from \oitrip.

A complementary comparison of spectral properties is presented in Fig.~\ref{fig_he_4epochs} where we compare all models (He-star progenitors evolved with a nominal mass loss) at 100, 200, 320, and 430\,d after explosion. In the next sections, we discuss in turn the strongest lines that represent landmarks in at least one model in our set. We start with \niidoub\ (Sect.~\ref{sect_nii}), followed by \oidoub\ and \mgi\ (Sect.~\ref{sect_oi}), and finally Ca\two, Ni\two\ and Fe\two\ lines (Sect.~\ref{sect_ige}).

\begin{figure}[h!]
\centering
\includegraphics[width=\hsize]{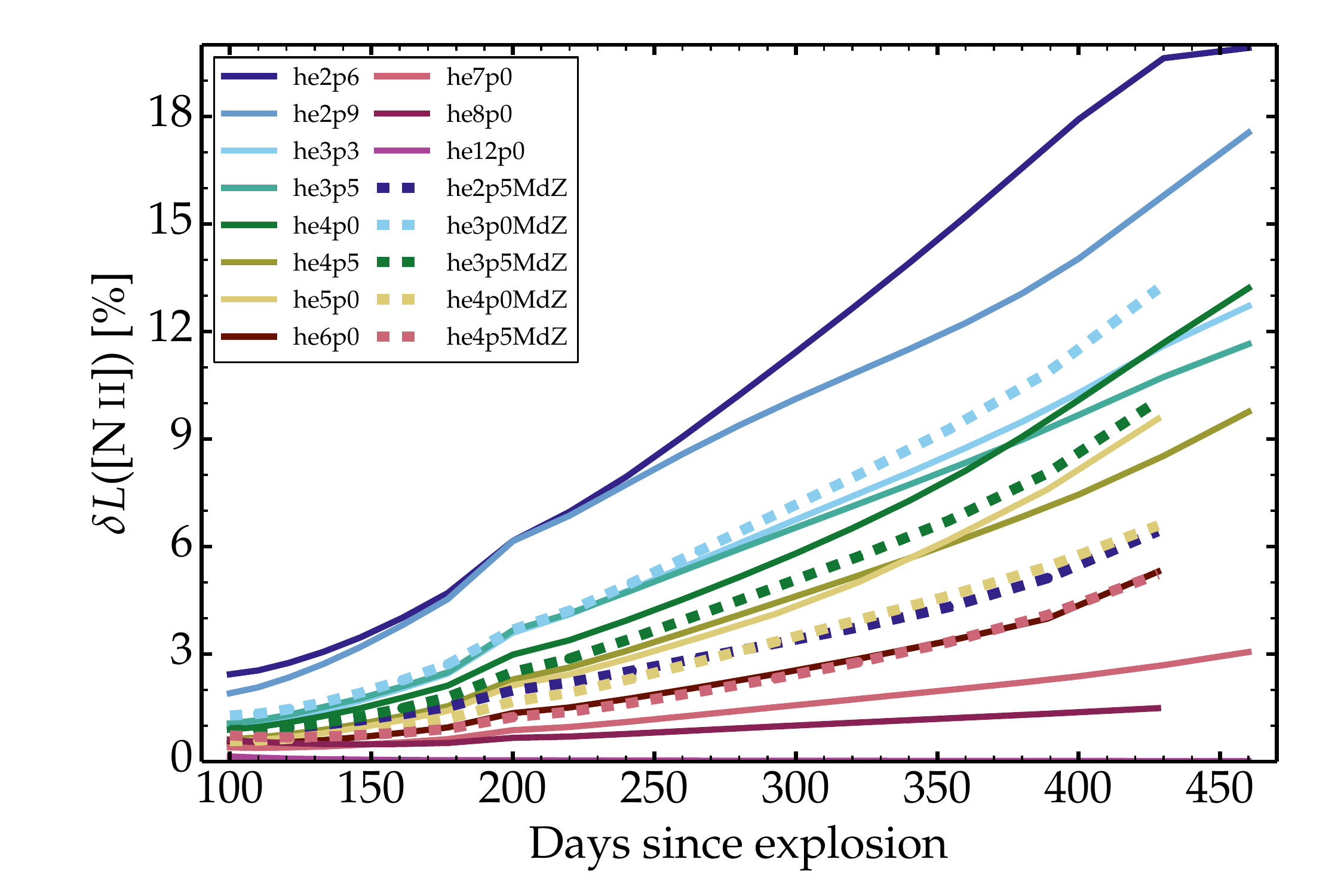}
\includegraphics[width=\hsize]{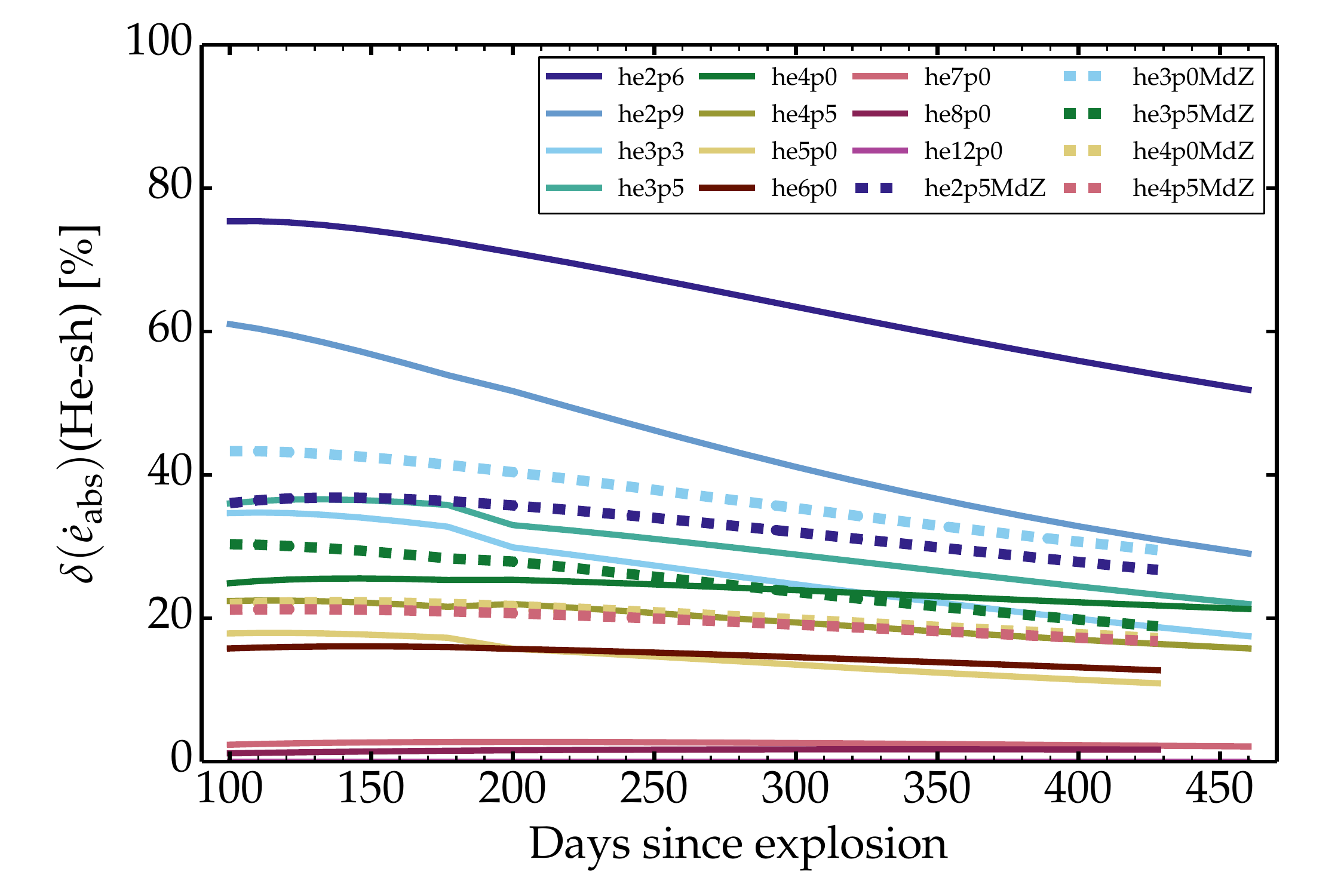}
\caption{
Top: Fractional line flux in \niidoub\ relative to the total optical luminosity versus time since explosion for our radiative-transfer calculations based on the ejecta models he2p6 to he12p0 arising from He-star progenitors evolved with the nominal mass loss rate \citep[solid line]{woosley_he_19,ertl_ibc_20} and on the ejecta models he2p5MdZ to he4p5MdZ arising from He-star progenitors evolved with no mass loss (dashed line). Bottom: Same as above but showing the fraction of the total decay power absorbed in the He-rich shell. Both are given as a percentage.
\label{fig_line_nii}
}
\end{figure}

\section{The He-rich shell and \niidoub}
\label{sect_nii}

Figure~\ref{fig_line_nii} shows the percentage of the optical flux (i.e., within 3500 and 9500\,\AA) radiated in \niidoub\ as a function of time for the full set of models. The integration of the line flux extends out to $\pm$\,12000\,\kms\ from the mean rest wavelength of the doublet (taken at 6565.5\,\AA) and hence the reported flux accounts for the contribution of both components. Because this doublet forms in the He/N shell, we also show the evolution of the power absorbed in the He-rich shell as a function of time (bottom panel).

This fractional flux is only a few percent at 100\,d but grows to 2--20\% at 450\,d. This increase is caused by various factors. The density drops as $1/t^3$ (where $t$ is the elapsed time since explosion) and eventually favors cooling through forbidden lines. N is primarily once ionized, favoring a high population of the N\two\ ground state (see Figs.~\ref{fig_ionization_evol}--\ref{fig_temp_evol}
 for additional information on the ionization and temperature in all ejecta shells). Finally,  the \niidoub\ doublet is eventually the dominant coolant for the He/N shell, swamping the cooling due to Fe\two. The strength of \niidoub\ varies between models because of the difference in decay power absorbed in that shell, which itself scales with the He/N shell mass. Model he2p6 (he8p0) is endowed with a He-rich shell that represents 92\% (15\%) of the ejecta mass and absorbs a fraction of the total decay power that is 75\% (1\%) at 100\,d but only 52\% (2\%) at 450\,d.

The temporal evolution of the decay power absorbed in the He-rich shell does not just depend on the mass of the He-rich shell. In model he2p6, the $\gamma$-ray escape from the ejecta is large so positron deposition increasingly dominates the total decay power absorbed (it represents 37\% at 450\,d in this model).  This power provides little benefit to the He-rich shell because it is \nifs\ free, and hence the fraction of energy emitted by the He-rich shell declines with time. However, \niidoub\ remains prominent since it radiates a larger fraction of the energy deposited in the He-rich shell. In higher mass models, the $\gamma$-ray trapping is greater so positron power plays a weaker role (see also Fig.~\ref{fig_edep_evol}). In that case, with time passing, $\gamma$ rays can travel some distance before being absorbed or escaping. In models more massive than he4p0, the power absorbed in the He-rich shell stays essentially constant with time.

\begin{figure}[h!]
\centering
\includegraphics[width=\hsize]{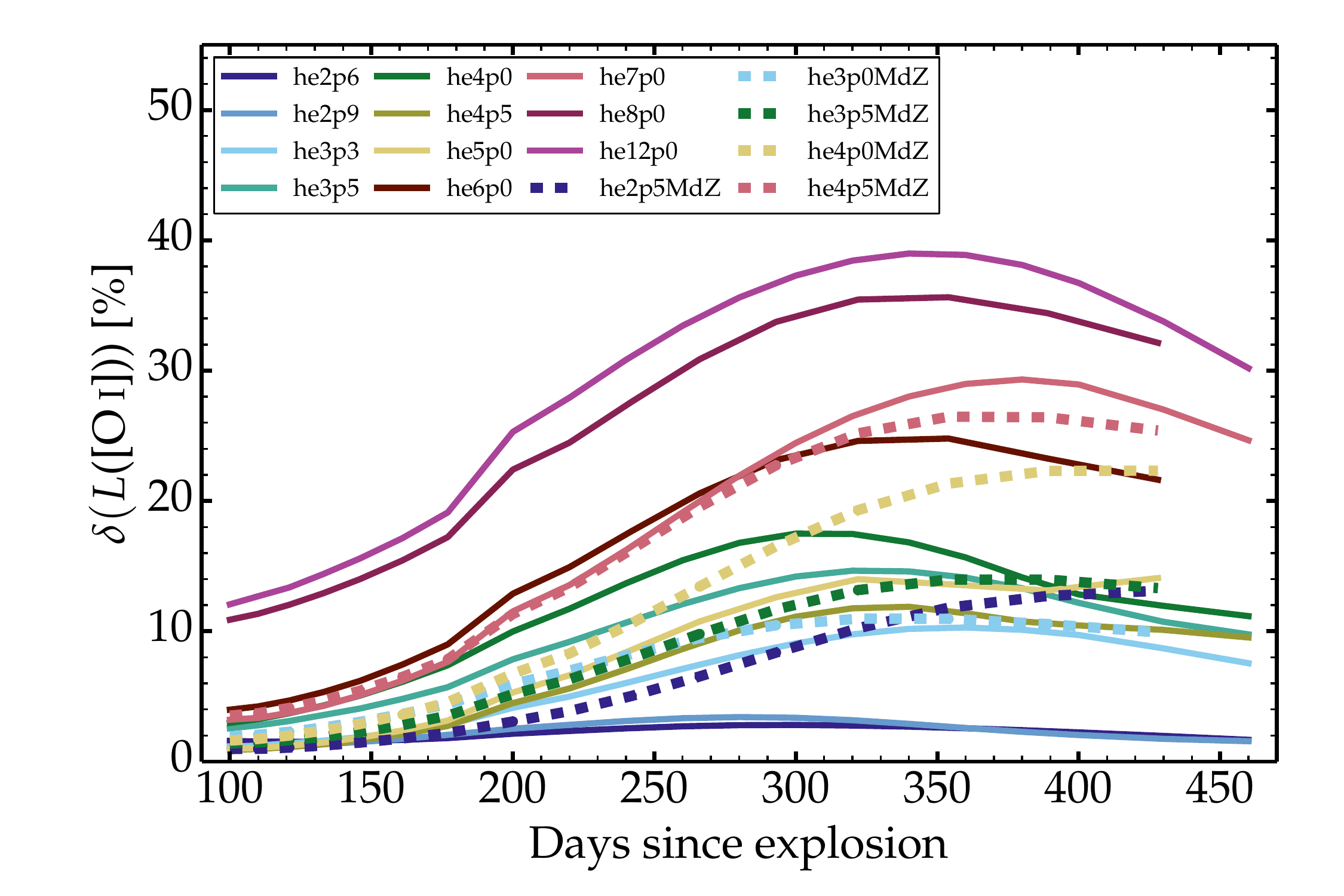}
\includegraphics[width=\hsize]{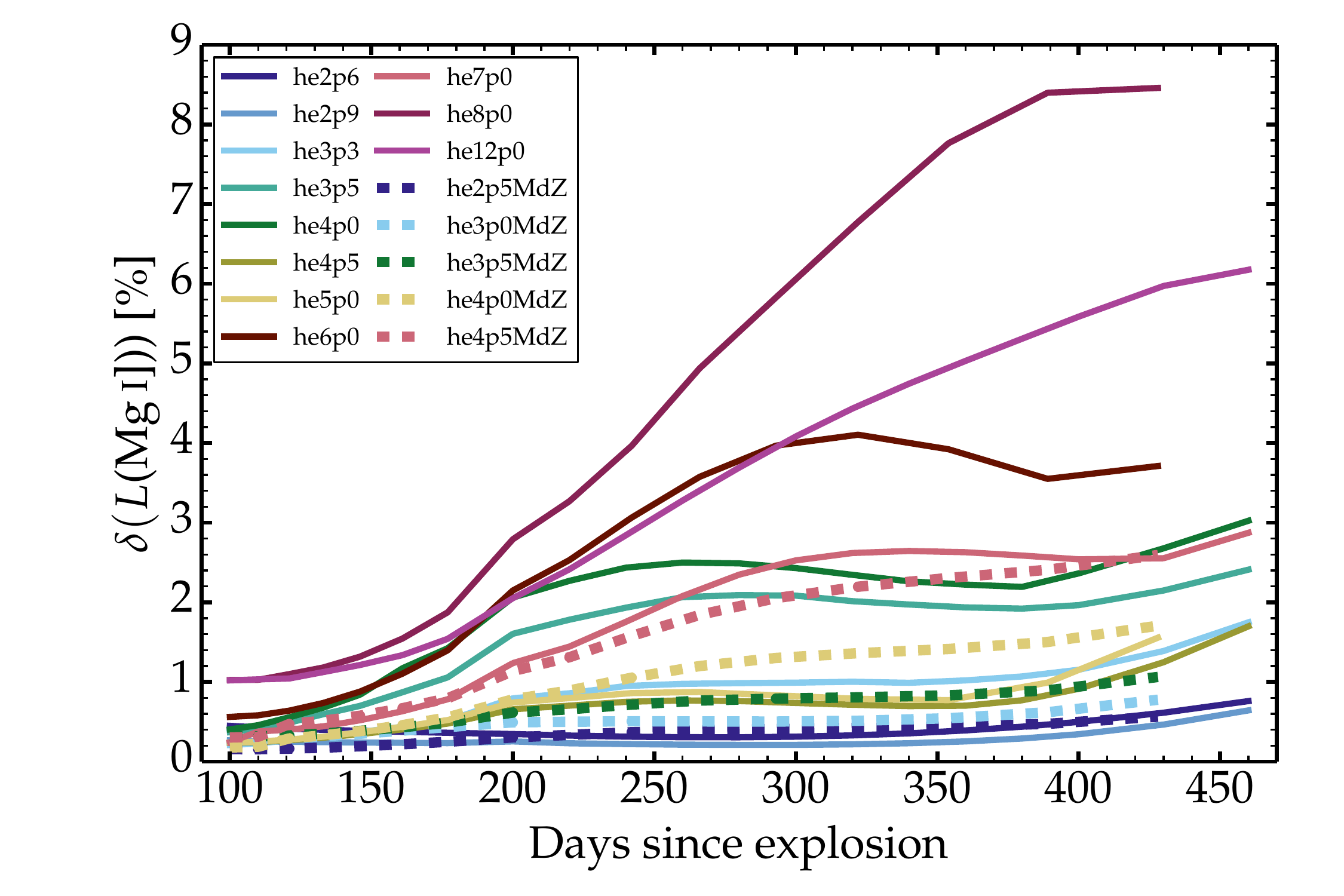}
\includegraphics[width=\hsize]{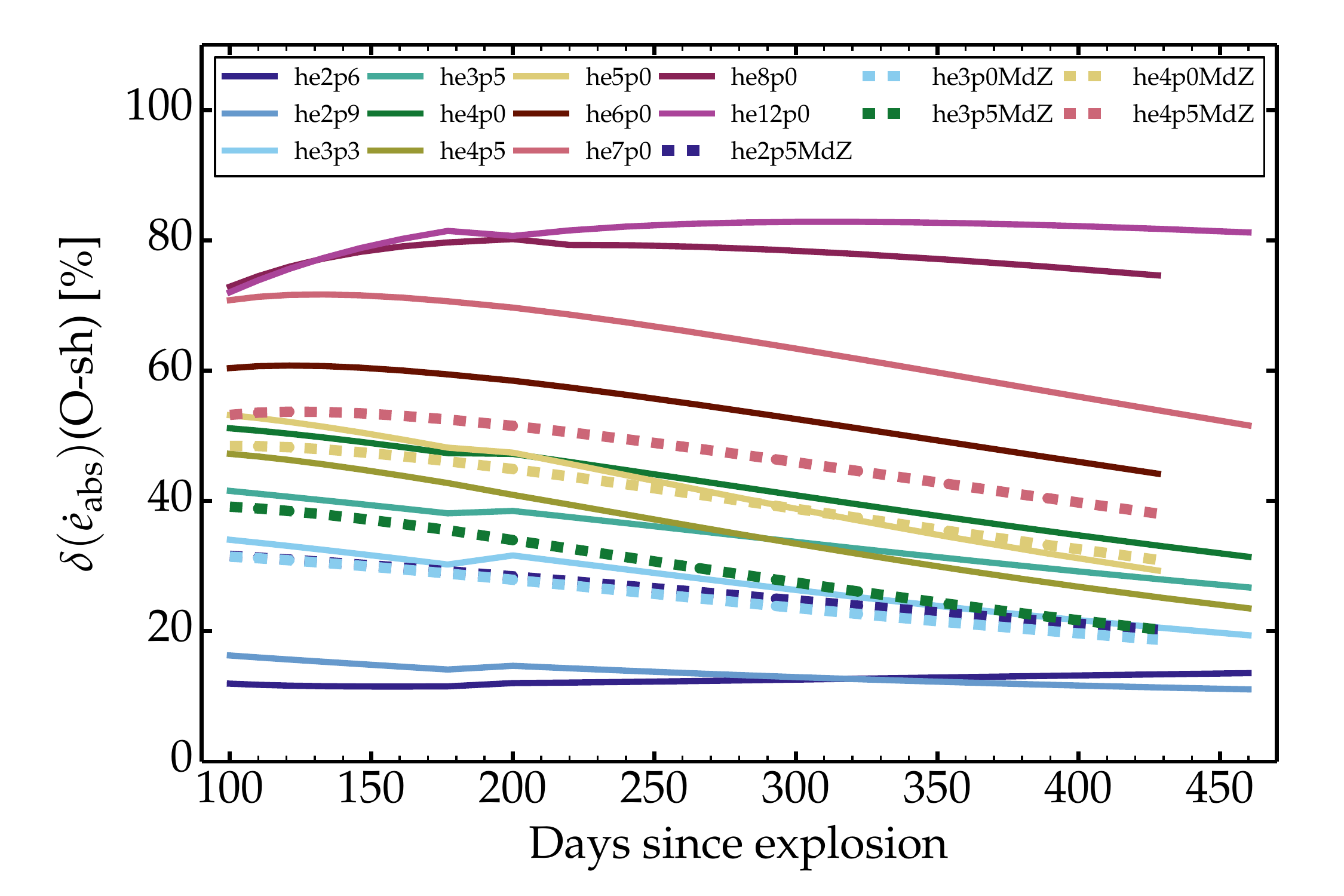}
\caption{Same as Fig.~\ref{fig_line_nii}, but showing the fractional flux associated with \oidoub\ (top panel) and \mgi\ (middle panel) as well as the fractional decay power absorbed in the O-rich shell (bottom panel).
\label{fig_line_oi}
}
\end{figure}

\section{The O-rich shell and \oidoub}
\label{sect_oi}

Figure~\ref{fig_line_oi} shows the percentage of the optical flux radiated in \oidoub\ (top panel) and \mgi\ (middle panel), as well as the fraction of the total decay power absorbed in the O-rich shell (bottom panel) as a function of time for all He-star explosion models (progenitors evolved with and without mass loss). Both fractional line fluxes increase with time but they seem to reach a maximum at about 300\,d for \oidoub\ and 400\,d for \mgi. Instead, the decay power absorbed in the O-rich shell decreases steadily in all models at all times except for models he8p0 and he12p0 for which there is first a rise from 100 to 200\,d and then a slow decrease.

In lower mass ejecta, the lower O yield and the lower O-rich shell mass lead to a weaker \oidoub. The progressive recombination of O contributes to strengthening \oidoub\ with time. In higher mass models, the O ionization is always low so recombination in time mostly contributes to decreasing the \oidoub\ and strengthening \mgi\ at times past 300\,d. In model he12p0, the \oidoub\ flux represents 10\% of the optical flux at 100\,d but 40\% at 350\,d. In low mass models, the \oidoub\ flux is just a few percent of the optical flux and always remains subdominant. The evolution of \mgi\ is qualitatively similar to that of \oidoub\ but shifted to smaller line fluxes because of the lower Mg abundance (see Table~\ref{tab_prog}). The delayed rise and peak stems from an ionization effect since Mg stays partially ionized for longer (the ionization potential of Mg\one\ is 7.6\,eV while that of O\one\ is 13.6\,eV). A similar effect occurs for Na\one, which also has a small ionization potential (i.e., 5.1\,eV) and \nad\ strengthens only later on. In model he12p0 (or he8p0), the drop in \oidoub\ flux after 350\,d is due to an increased leakage of power into \mgi\ and \nad.

In high mass progenitors, the O-rich shell represents the bulk of the ejecta mass (97\% in model he12p0) and captures the bulk of the decay power. The \oidoub\ is not strong early on because the density is too high (which inhibits forbidden line formation) -- there is little change in O ionization in the O-rich shell in that model. In lower mass models, the O-rich shell is smaller (it represents 45\% of the total mass in model he5p0) and absorbs less power but the O ionization is partial so a fraction of that power is radiated by Fe\two\ lines. Both effects conspire to reduce the \oidoub\ line strength and benefit Fe\two\ line emission, particularly at early times. But the progressive cooling leads to the strengthening of both \oidoub\ and \mgi.

Clumping can impact  the strength of \oidoub\ in models that exhibit at any epoch a partial O ionization (D21). Clumping does not boost the strength of the emission directly, but instead indirectly through the recombination  it induces. D21 found that a modest compression of the O-rich material can enhance the population of neutral O, which eventually takes place in most models at 300--400\,d.  Uncertainties in clumping are thus less impactful at late times, making them more suitable to infer the yields from SNe Ib and Ic in that respect. Unfortunately, molecule and dust are likely to start forming at these times (see, for example, \citealt{jerkstrand_15_iib}; \citealt{rho_20oi_21}). Since the effects of clumping have been documented in detail in D21, they are not repeated here.

\begin{figure*}[h!]
\centering
\includegraphics[width=0.495\hsize]{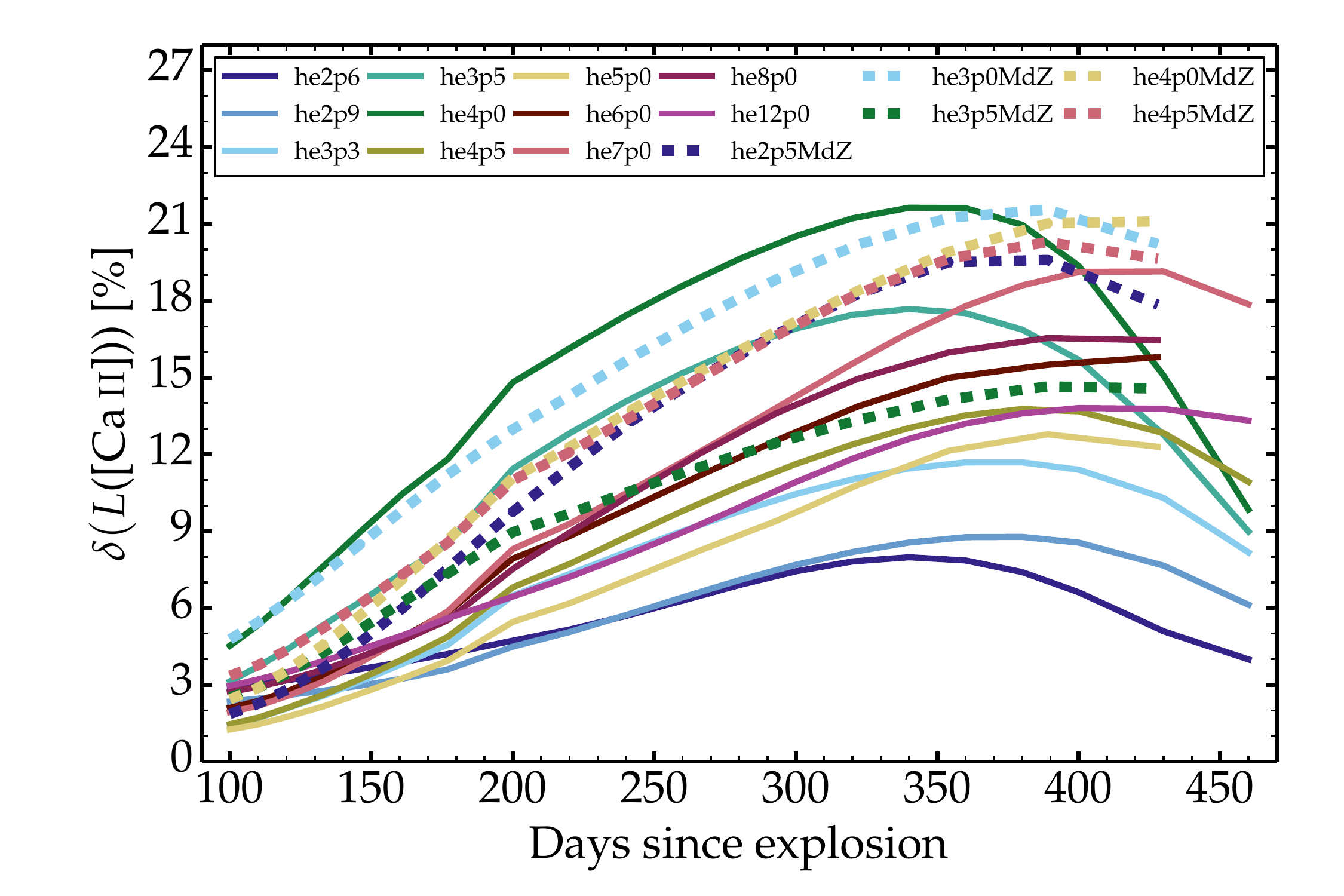}
\includegraphics[width=0.495\hsize]{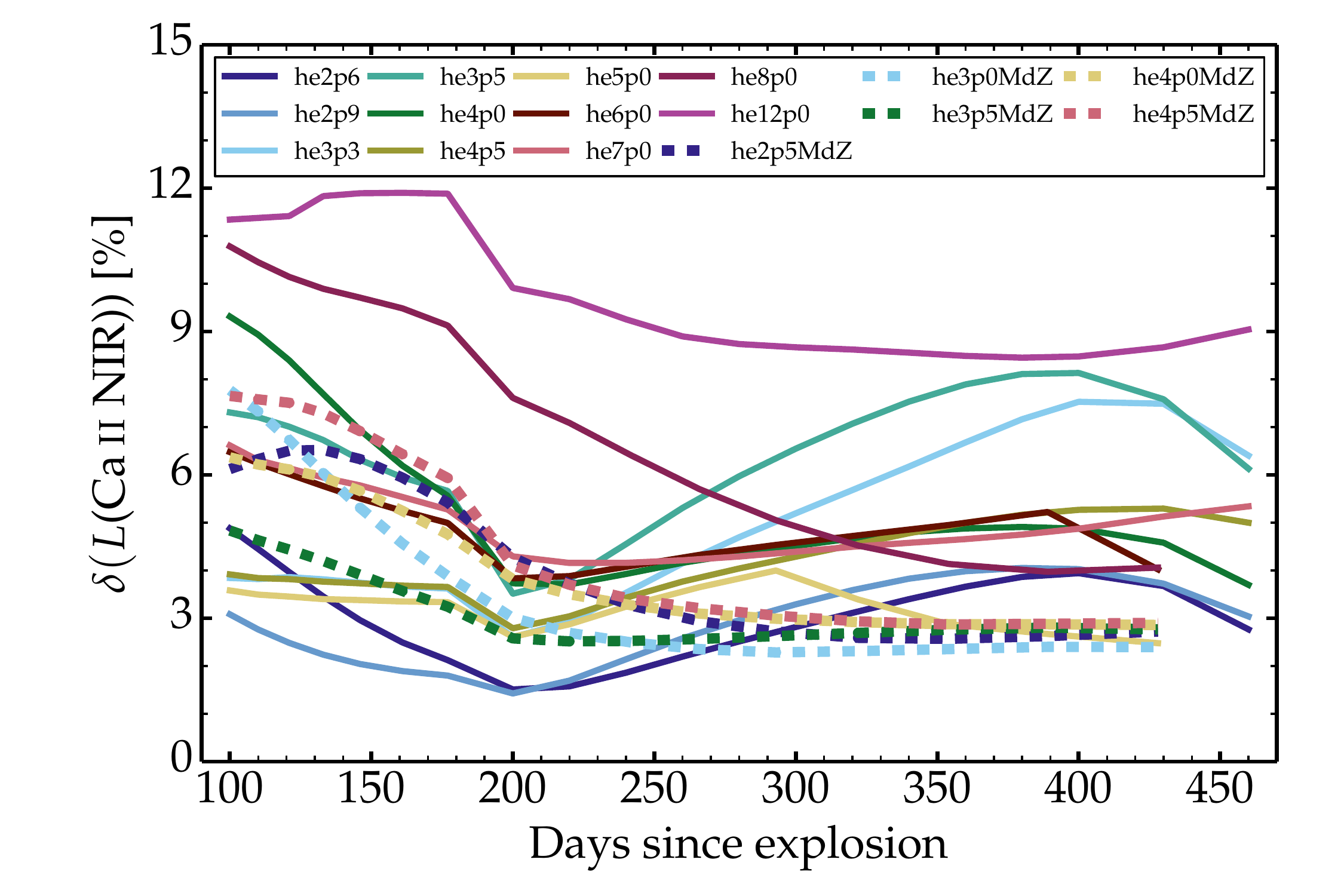}
\includegraphics[width=0.495\hsize]{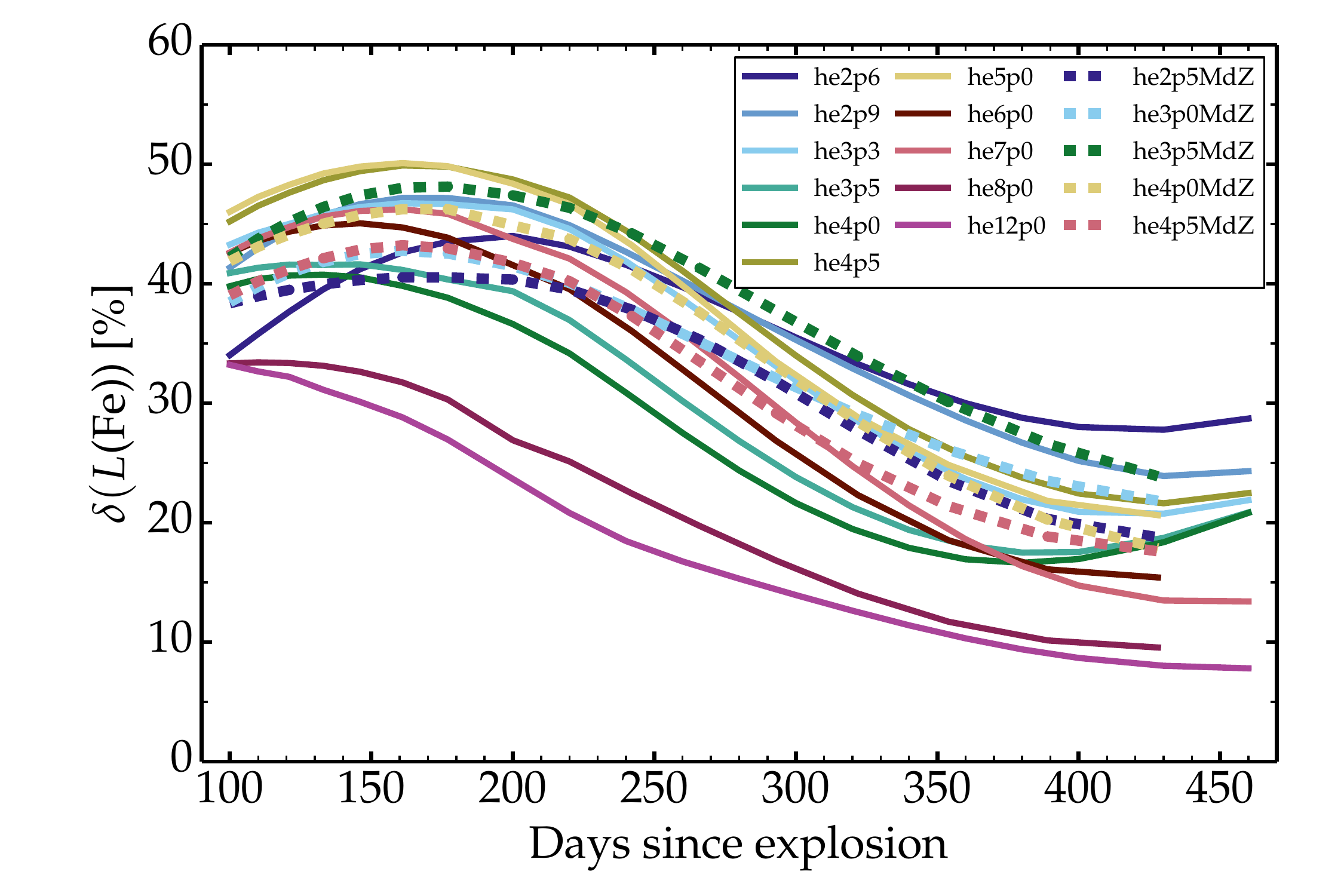}
\includegraphics[width=0.495\hsize]{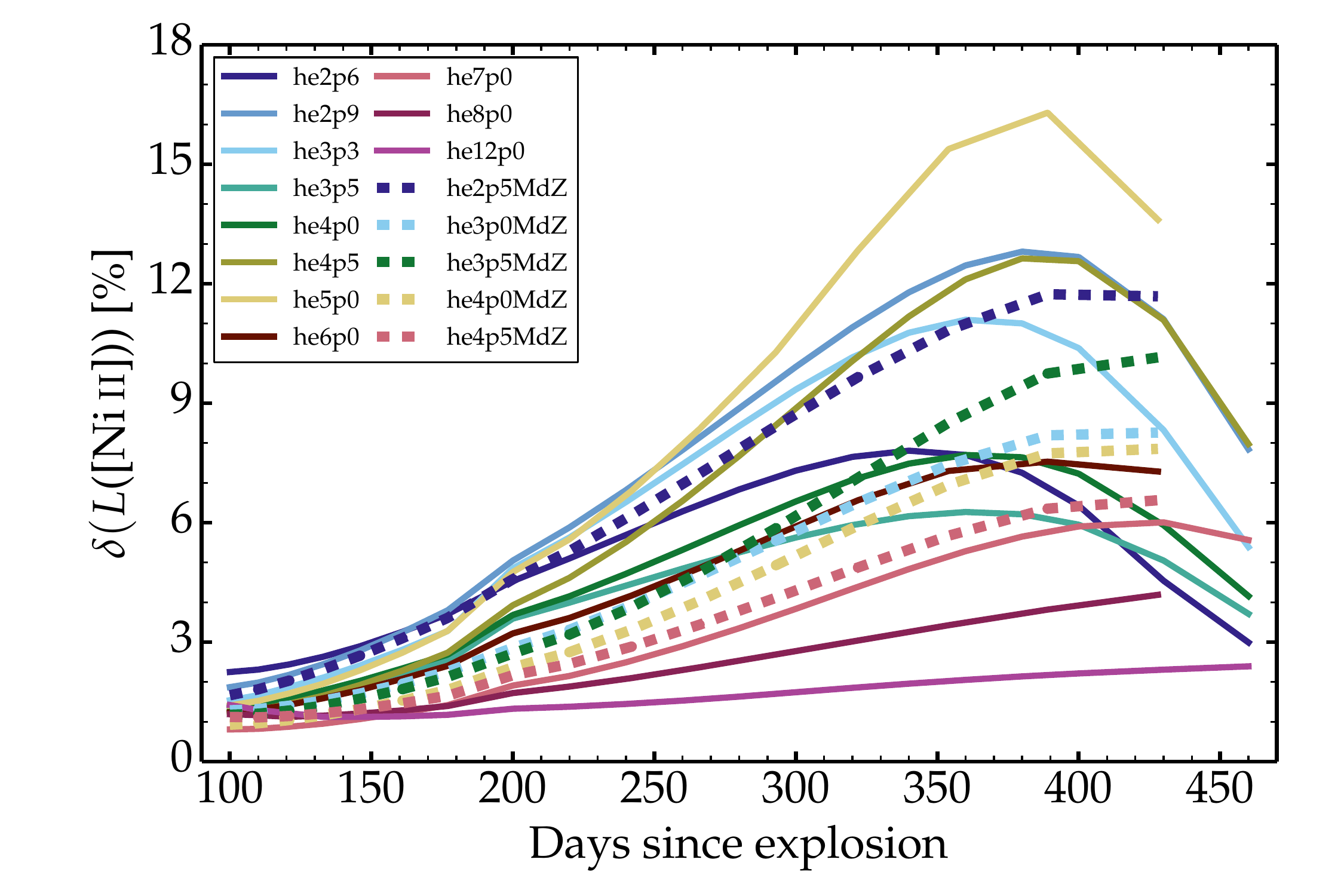}
\includegraphics[width=0.495\hsize]{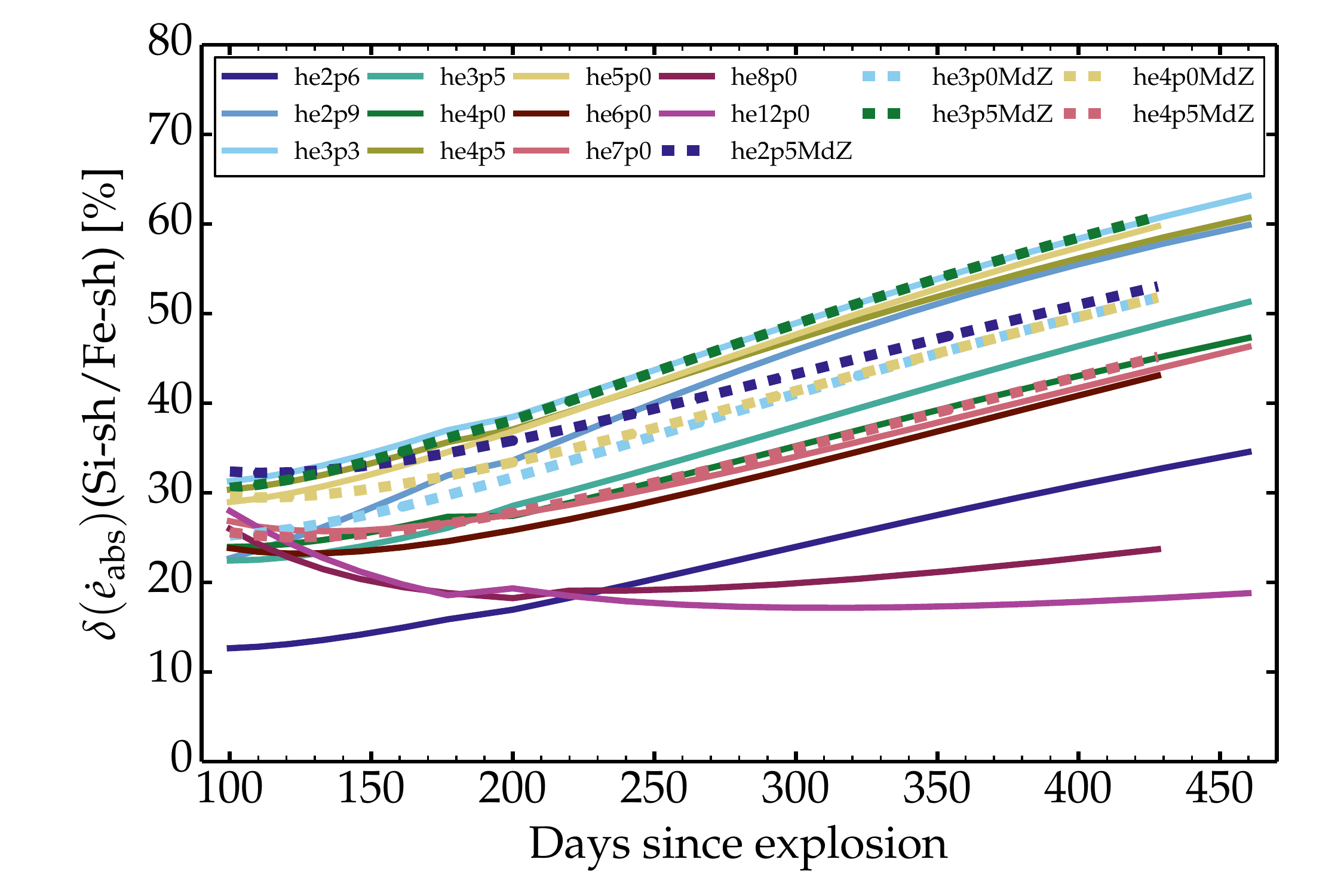}
\caption{Same as Fig.~\ref{fig_line_nii}, but showing the fractional flux associated with \caiidoub\ (top left  panel), with \caiitrip\ (top right panel), with  Fe emission in the 4100-5500\,\AA\ range (middle left panel), and with [Ni\two]\,$\lambda$\,$7378$ (middle right panel), as well as the fractional decay power absorbed in the combined Si-rich and Fe-rich shells (bottom panel).
\label{fig_line_si_fe}
}
\end{figure*}

\section{The Fe/He and Si/S shells and Ca\two, Ni\two, and Fe\two\ lines}
\label{sect_ige}

Figure~\ref{fig_line_si_fe} shows some important diagnostics for the Si-rich and Fe-rich shells. Both shells were grouped here for simplicity because they are thin and hard to distinguish in our models. Furthermore, their composition is in fact similar in many ways, with Ca and \nifs\ abundant in both shells (D21). Figure~\ref{fig_line_si_fe} shows the evolution of the fraction of the optical flux emerging in \caiidoub\ (top left panel), in \caiitrip\ (top right panel), in \nkiiopt\ (middle left panel), in the iron lines (middle right panel), as well as the fraction of the total decay power absorbed in the combined Si-rich and Fe-rich shells (bottom row panel), for all He-star explosion models (progenitors evolved with and without mass loss). While Ca is most abundant in the Si-rich shell, the \caiidoub\ forms both in the Fe-rich and the Si-rich shells and the \caiitrip\ may form everywhere, including in the O-rich shell where its abundance is subsolar (see D21 for discussion).

Because the Ca-rich layers are formed through explosive burning, the Ca abundance scales with the \nifs\ mass and is thus not an ideal diagnostic of the progenitor mass (unlike O).  In our set of models, the \nifs\ mass is between 0.01 and 0.1\,\msun, with a lot of scatter between models although there is a tendency for small values at the low mass end (the minimum \nifs\ mass of 0.01\,\msun\ is for the lightest model in the sample, i.e., he2p6).  The \caiidoub\ doublet line strength is seen to rise in all models until about 350\,d. This rise stems in part from the drop in density and in ionization, and mirrors the drop in iron line emission. In lower-mass models, this rise stems also from the fact that the decay power absorbed in the combined Si-rich and Fe-rich shells increases in time following the growing fraction of local energy deposition from positrons while $\gamma$ rays increasingly escape the ejecta (and deprive the O-rich and He-rich shells).

The apparent strength of the 7300\,\AA\ feature in our spectra comes in part from \nkiiopt. Stable Ni (i.e., \nife) is present with a mass between about 0.001 and 0.008\,\msun. In the model most abundant in \nife\ (namely he5p0), the \nkiiopt\ line flux represents $\sim$\,1\% of the total optical flux at 100\,d but this fraction rises to $\sim$\,15\% at 400\,d. Not knowing a priori the abundance of \nife\ in any observed SN (in the infrared, the less blended [Ni\two] lines at 1.94 and 6.63\,$\mu$m might help reduce this uncertainty), it is unclear what fraction of the 7300\,\AA\ line flux is due to \caiidoub. This emission at 7300\,\AA\ should therefore be used with caution in any mass estimate.

The iron emission in the range 4000 to 5500\,\AA\ represents a significant fraction of the SN radiation at 100\,d in all models, and this persists at all times covered here in lower mass models. Fe\two\ emission occurs not just in the Fe/He shell but throughout the ejecta. Partial ionization at early times (which prevents strong forbidden lines like \oidoub\ to form) favor Fe\two\ cooling in the O-rich shell. In lower mass He-star models, Fe\two\ is an important coolant of He-rich material (in addition to N\two). In the O-rich shell, Fe\two\ cooling tends to persist as long as ejecta densities are too high for the formation of forbidden lines (these appear when radiative de-excitation is more likely than collisional de-excitation).

\section{Discussion and conclusions}
\label{sect_conc}

We have presented a grid of time-dependent nonLTE radiative transfer simulations computed with \cmfgen, covering the nebular phase from 100 until about 450\,d after explosion, and based on He-star progenitor and explosion models originally computed by \citet{woosley_he_19} and \citet{ertl_ibc_20}. This study is an extension of our previous work (D21) on similar models but in which we focused on the properties  at a single epoch of 200\,d after explosion. These calculations currently represent the largest grid of radiative transfer models for SNe Ibc at nebular epochs. Despite the large number of models, we ignored the possibility of clumping, molecule formation, or departures from spherical symmetry so that this model grid underestimates the real extent of the parameter space covered by SNe Ib and Ic as well as the degeneracy of their radiative properties. To remain concise, this presentation was limited to model results. Details about the underlying ejecta properties or the complications of line formation in SNe Ibc were presented in D21. A comparison of the present models, together with those discussed in D21, with the observations of Type Ib and Ic SNe will be presented  elsewhere.

As discussed in D21, the preSN yields of He-star models depend on preSN mass loss such that He stars dying with the same final mass have a comparable O yield (or CO core mass). This implies that evaluating the nucleosynthesis of SN Ibc ejecta can at best constrain the preSN mass but not the original mass on the main sequence. It also implies that Type Ibc progenitors of a given mass on the main sequence will tend to contribute less metals at higher metallicity. To address this aspect and quantify its impact, we computed a new set of models in which no preSN mass loss was assumed. While our He-star models likely result from the early removal of the H-rich envelope in an interacting binary at the onset of He-core burning, these zero mass loss models correspond to a similar removal taking place just before core collapse. The two sets of models probably bracket the range of possibilities for the timing of the envelope removal as well as uncertainties in wind mass loss rate. To keep in line with the low ejecta masses inferred from observations of SNe Ibc, we limit this set of zero mass loss models to low initial He-star masses (i.e., 2.5 to 4.5\,\msun), yielding ejecta masses in the range 1.1--2.8\,\msun.

In this comparison, we showed that models he4p5MdZ and he6p0, which die with a final mass of 4.5 and 4.44\,\msun, and He mass of 1.15 and 0.95\,\msun, and an O mass of 0.97 and 0.96\,\msun,  produce similar spectra over the 100 to 450\,d timespan -- models with the same final, preSN mass have similar yields and produce essentially the same emergent spectra. Only weak and transitory  differences are apparent. He\one\,7065\,\AA\ is stronger prior to $\sim$\,150\,d  in model he4p5MdZ. The 40\% greater \nifs\ mass in that model yields a greater brightness and is also likely at the origin of the slightly stronger \caiidoub\  since that line forms in the Fe/He and Si/S shell, which are both explosively produced. Numerous finer differences in these two ejecta and progenitor models, including different O/Ne/Mg and He/C shell masses, the C and O mass fraction in the He/C shell, or the different $E_{\rm kin}/M_{\rm ej}$, are hard to diagnose and suggest further degeneracy of nebular-phase spectra. The uncertain progenitor mass loss history thus compromises the inference of the initial, main sequence mass of SNe Ib and Ic -- this shortcoming does not affect SNe IIb whose progenitor He core is intact at the time of core collapse.

Our broad set of models from low-mass to high-mass He star progenitors yield SN Ibc spectra with large differences at all epochs. That is, models that differ significantly in preSN mass have widely different spectral properties at any given time of the nebular evolution. These differences reflect the different masses of the Fe/He, Si/S, O-rich (i.e., O/Si, O/Ne/Mg and O/C), and He-rich (i.e., He/C and He/N) shells and the fact that the power absorbed in a shell scales to first order with its mass. Hence, at the low He-star mass end, the preSN models have a composition dominated by the He-rich shell with small metal yields. Such models exhibit strong Fe\two\ emission in the 4000-5500\,\AA\ range (this emission is always stronger early on) and weak \oidoub\ at all times. He\one\ lines are present up until about  150\,d, while N\two\ lines are present at all nebular epochs, with \niidoub\ even becoming the strongest optical line at 450\,d.

Higher mass models, characterized by a decreasing He-rich shell mass and an increasing O-rich shell mass, show weaker emission from the He-rich shell (weak or no He\one\  and N\two\ lines) while the emission from the O-rich shell strengthens. At early times, and in the absence of clumping, O is partially ionized in lower mass He-star models so a fraction of the O-rich shell cools through Fe\two\ emission. However, after about 300\,d, O is essentially neutral in the O-rich shell and the \oidoub\ is strong, and the more so for larger He-star models. The \caiidoub\ emission is a poor tracer of the preSN mass because its strength scales with the Fe/He and Si/S shell masses, which result from explosive nucleosynthesis.

Quantifying the nucleosynthetic yields and characterizing the progenitors of Type Ibc  SNe from nebular phase spectra seems easier at times later than 300\,d. At such late times, the ejecta cool through a few strong lines rather than a myriad of lines. Furthermore, the ionization in the O-rich shell is predicted to be low (O is essentially neutral), even in the absence of clumping. Hence, the uncertain level of clumping is less critical.  However, at these epochs cooling by CO and other molecules is likely to be important, and will need to be considered in future models.

More quantitative information can be gleaned by studying the evolution of line fluxes through the nebular phase (as done here from 100 to 450\,d) rather than focusing on one or two epochs. Indeed, the full evolution reveals the change in temperature or ionization of ejecta shells of different composition, the evolving power contribution of $\gamma$-rays and positrons etc. As we argue in \citet{D22_lsst}, the Vera C. Rubin Observatory Legacy Survey of Space and Time will provide important information on the photometric evolution of a large sample of SNe Ibc, which will complement what has been achieved so far primarily with spectroscopy.

Inferring the mass of \nifs\ from SN Ibc nebular spectra is challenging. This may be done in SNe Ia because the \nifs\ mass is typically half the ejecta mass, the ionization is relatively high allowing one to observe lines of Fe\two--\three\ and Co\two\--\three\ simultaneously and track the evolution of their relative strength in time \citep{kuchner_co3_94,d14_co3}. In contrast, the mass of \nifs\ is only a few percent of the ejecta mass in SNe Ibc, so that at nebular times a significant part of the Fe\two\ emission is from primordial Fe (a \nifs\ mass analogous to that inferred in SNe Ia is unlikely since in that case the nebular spectra would exhibit strong emission lines from iron-group elements). There also seems to be no strong Co line in the optical to infer an Fe/Co abundance ratio. This requires further investigation.

\begin{figure}[t!]
\includegraphics[width=\hsize]{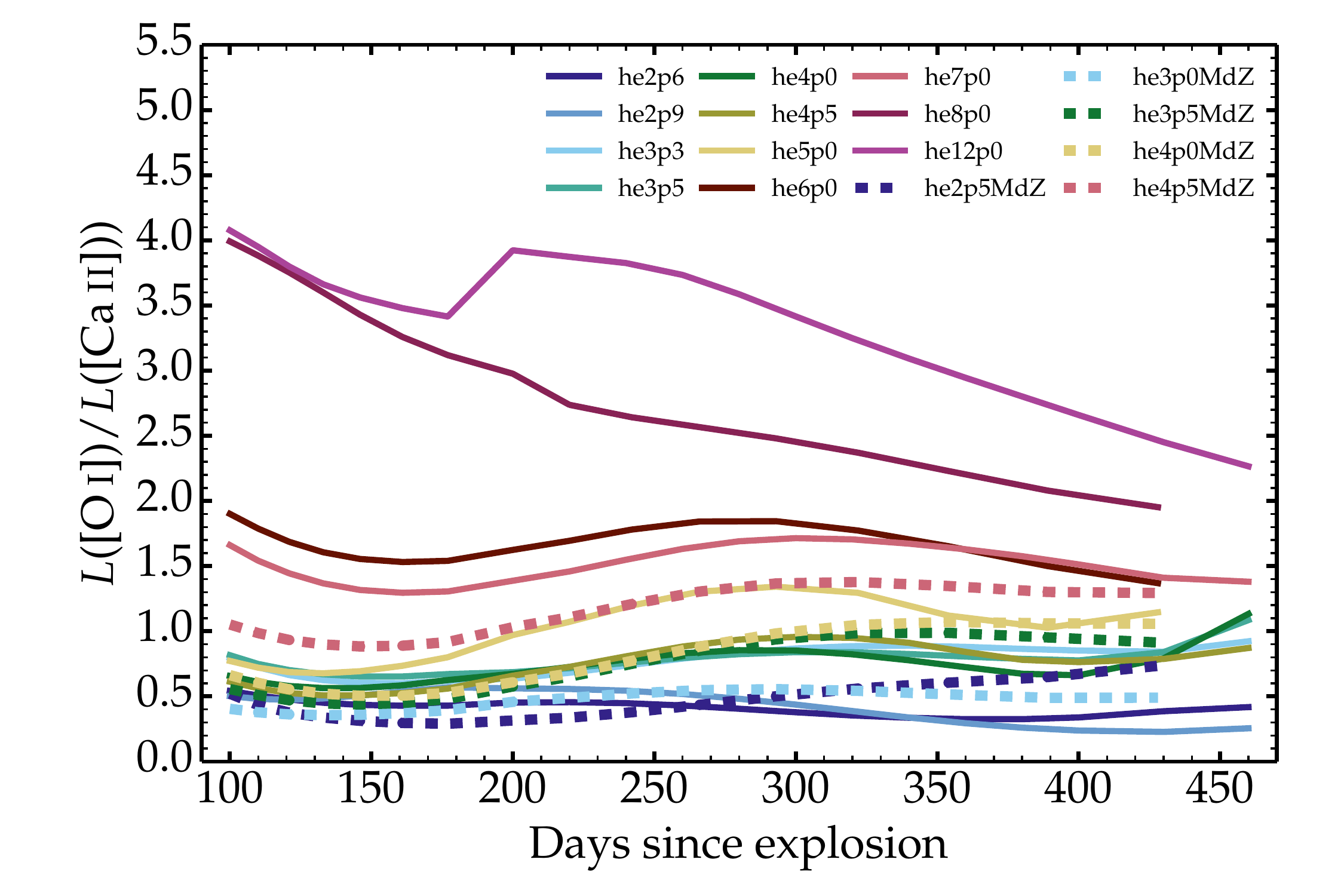}
\caption{Same as Fig.~\ref{fig_line_nii}, but showing the ratio of the \oidoub\ flux with the \caiidoub\ flux.
\label{fig_oi_over_caii}
}
\end{figure}

Models at the low mass end, whose composition is dominated by He (i.e., models lighter than he3p5, although He content is not the only characteristics influencing the observables), seem to be in conflict with the currently observed properties of Type Ib SNe. These models exhibit strong N\two\ lines and weak \oidoub, while Type Ib and Type Ic SNe tend to show similar nebular phase spectra, with a tendency for weaker \oidoub\ in some Type Ib SNe \citep{fang_neb_22}. Although unlikely, this discrepancy could be an observational bias resulting from the lower \nifs\ mass in lower mass He-star explosions making such events harder to detect, and even when found, less likely to be observed at nebular times. Another explanation  may be that lower mass He-star models become He giants at the end of their lives \citep{yoon_presn_12,eldridge_13bvn_15,clelland_he_16,woosley_he_19}, enhancing the likelihood for mass transfer to a companion or for a common envelope phase \citep{laplace_stripped_20} -- our He-star models were evolved in isolation from the He zero age main sequence until core collapse. In this case, such models could lead to ultra-stripped Type Ic SNe \citep{dewi_pols_03,tauris_ulstr_13} or to Type Ibn SNe \citep{pasto_ibn_08,hosseinzadeh_ibn_17,dessart_ibn_22}. This possibility requires further study.

Our grid of 1D models exhibits a clear correlation between preSN mass and the ratio of the \oidoub\ flux with the \caiidoub\ flux (Fig.~\ref{fig_oi_over_caii}). Such a correlation seems absent in observations \citep{fang_neb_22,prentice_neb_22}, although stripped-envelope SNe with weaker \oidoub\ tend to be of Type Ib \citep{fang_neb_22}. There may be various reasons for that. While the \oidoub\ flux arises from the O-rich shell, which is made prior to explosion, the \caiidoub\ flux arises from the Fe/He and Si/S shells, which are both created or reset during explosive nucleosynthesis. The flux in both lines depends on the \nifs\ mass, and the flux in the \oidoub\ depends also on the ionization and clumping in the O-rich shell (\citealt{jerkstrand_15_iib}; D21). The distinct [O\one]/[Ca\two] flux ratio between models is probably driven from the large range in O-shell masses in our model set, which supersedes other dependencies.

The ejecta with the weakest \oidoub\ line flux arise from the lower mass He-star models, which cool primarily through Fe\two\ and N\two\ line emission, and for which, as noted earlier, there are currently no observational counterparts. Higher-mass He-star models in our grid are not favored by the initial mass function and may thus be rare in the observed sample of SNe Ibc. Hence, the lack of correlation in the observed [O\one]/[Ca\two] flux ratio might indicate that the bulk of SNe Ibc are of comparable and moderate preSN mass (or at least that the current sample of SNe Ibc preSN masses covers a reduced range relative to our model set). It also suggests that what distinguishes the majority of SNe Ib and Ic may not be so much the composition (i.e., total yields) but the large scale mixing of \nifs\ and other elements like He and O \citep{d12_snibc}.  Further work is needed to quantify the role of large-scale mixing for the typing of stripped-envelope SNe.

\begin{acknowledgements}

This work was supported by the ``Programme National de Physique Stellaire'' of CNRS/INSU co-funded by CEA and CNES. DJH thanks NASA for partial support through the astrophysical theory grant 80NSSC20K0524. H.K. was funded by the Academy of Finland projects 324504 and 328898. This work was granted access to the HPC resources of  CINES under the allocation 2020 -- A0090410554 and of TGCC under the allocation 2021 -- A0110410554 made by GENCI, France. This research has made use of NASA's Astrophysics Data System Bibliographic Services.

\end{acknowledgements}


\begin{appendix}

\section{Additional information}

\begin{figure}
\centering
\includegraphics[width=\hsize]{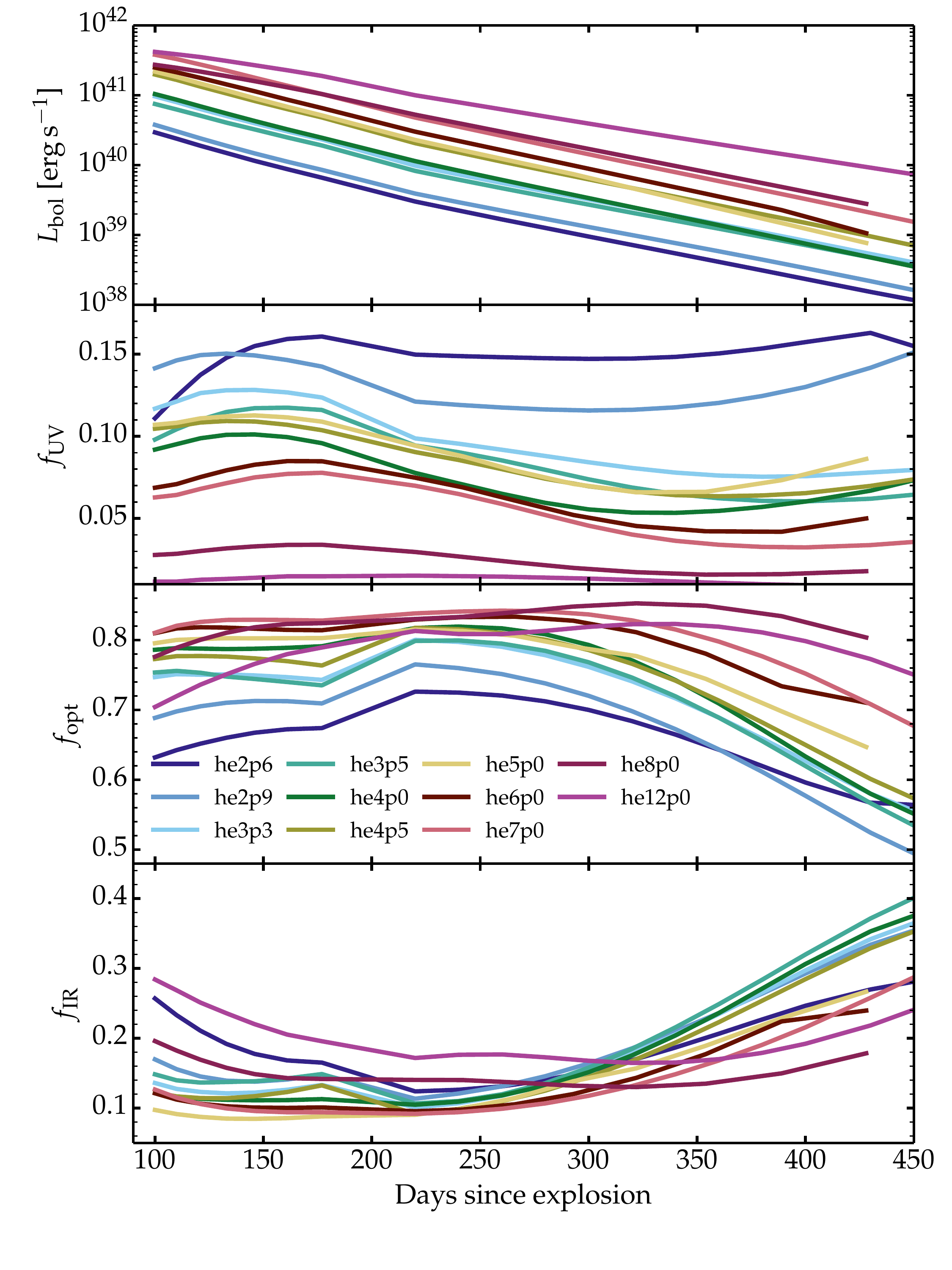}
\caption{Evolution of the bolometric luminosity and the fraction of that luminosity falling in the UV, the optical, and the IR ranges as a function of time for the He-star explosion models from progenitors evolved with a nominal mass loss (see \citealt{woosley_he_19}; \citealt{ertl_ibc_20}; D21).
\label{fig_lbol_frac}
}
\end{figure}

\begin{figure}
\centering
\includegraphics[width=\hsize]{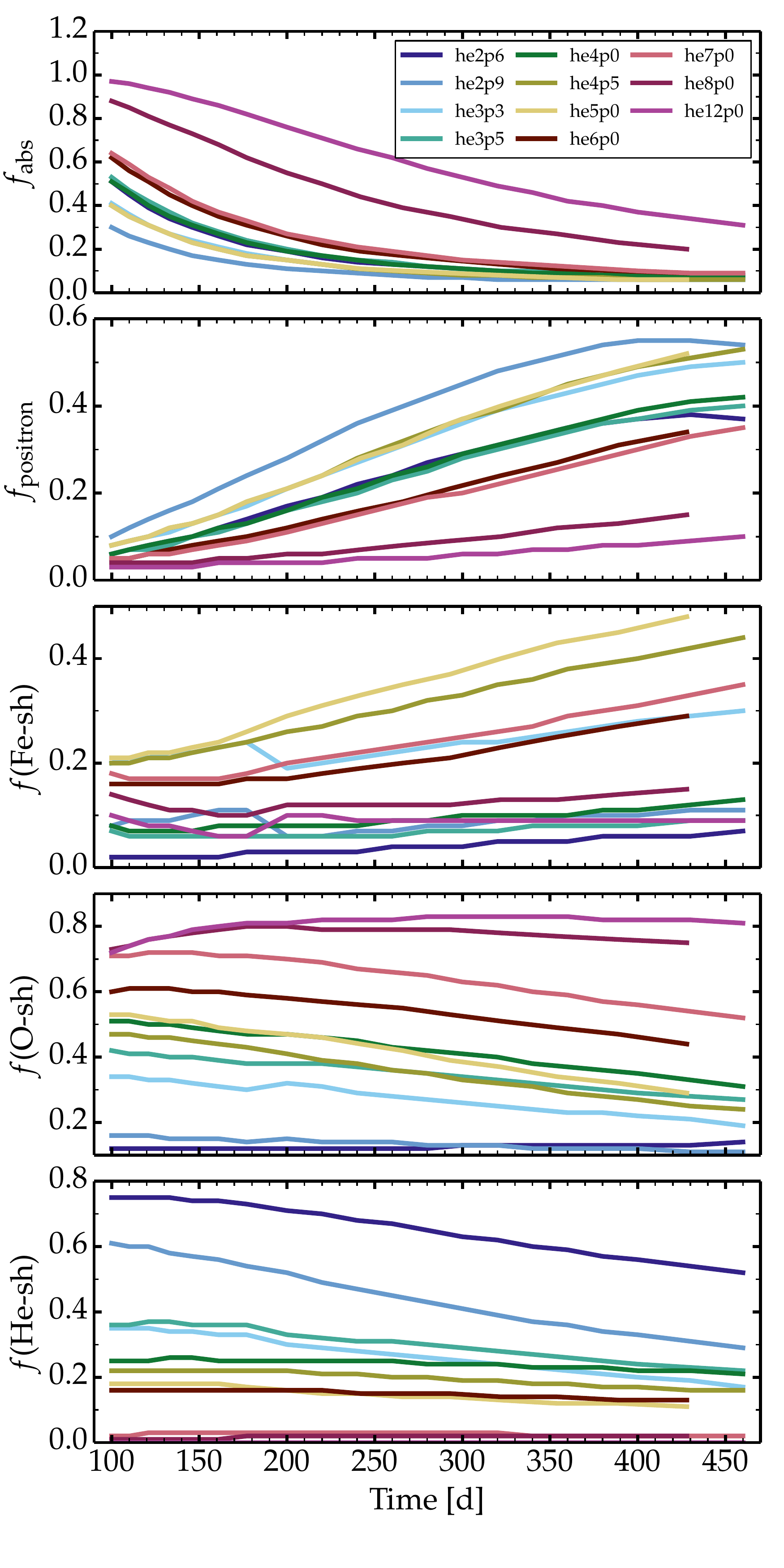}
\caption{Evolution of the decay power absorbed in the ejecta, the fraction of that power that arises from positrons (assumed here to be absorbed locally), and the fraction of the total decay power absorbed by the Fe-rich, O-rich, and He-rich shells in models he2p6 to he12p0.
\label{fig_edep_evol}
}
\end{figure}

\begin{figure*}
\centering
\includegraphics[width=0.45\hsize]{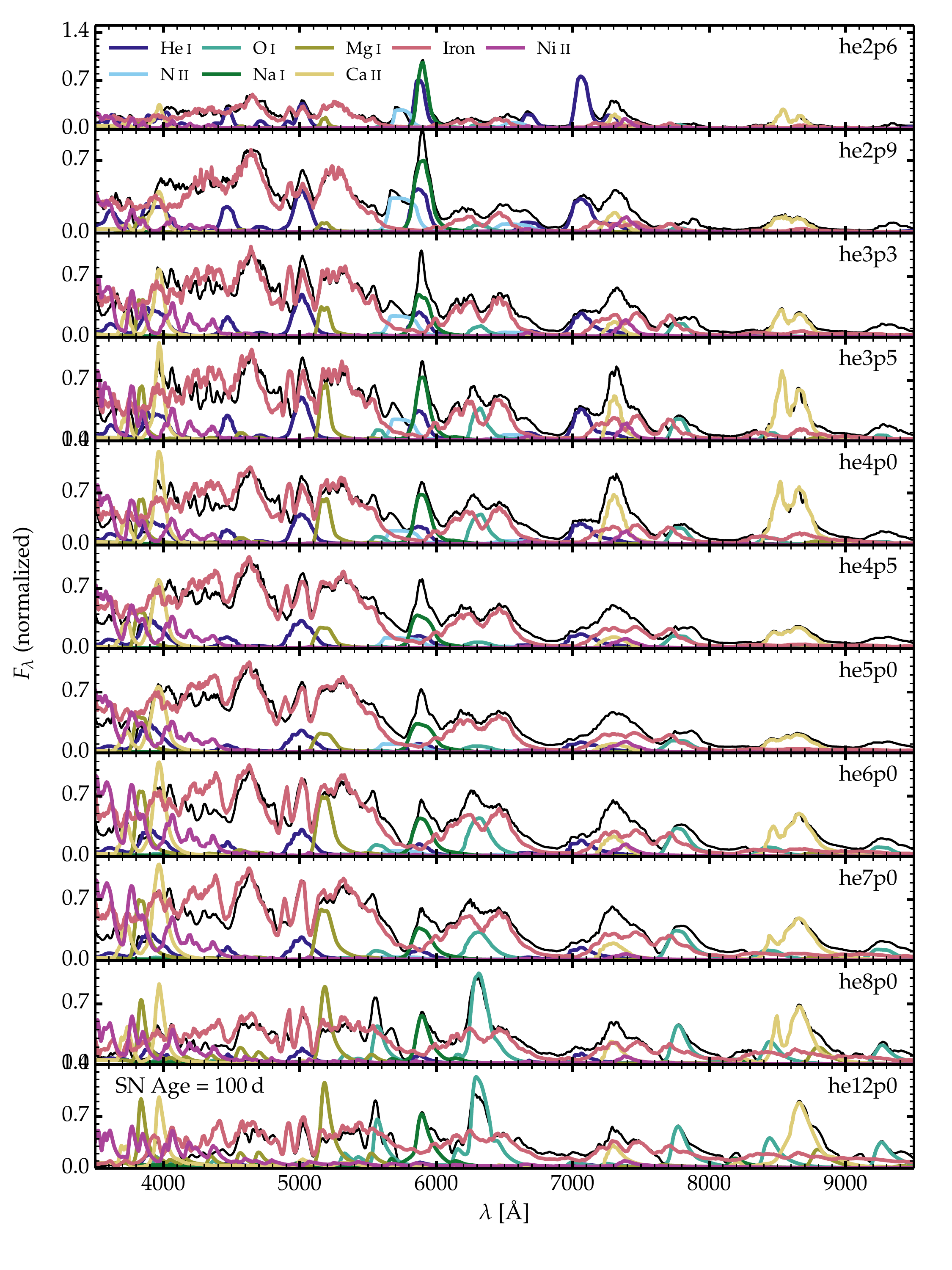}
\includegraphics[width=0.45\hsize]{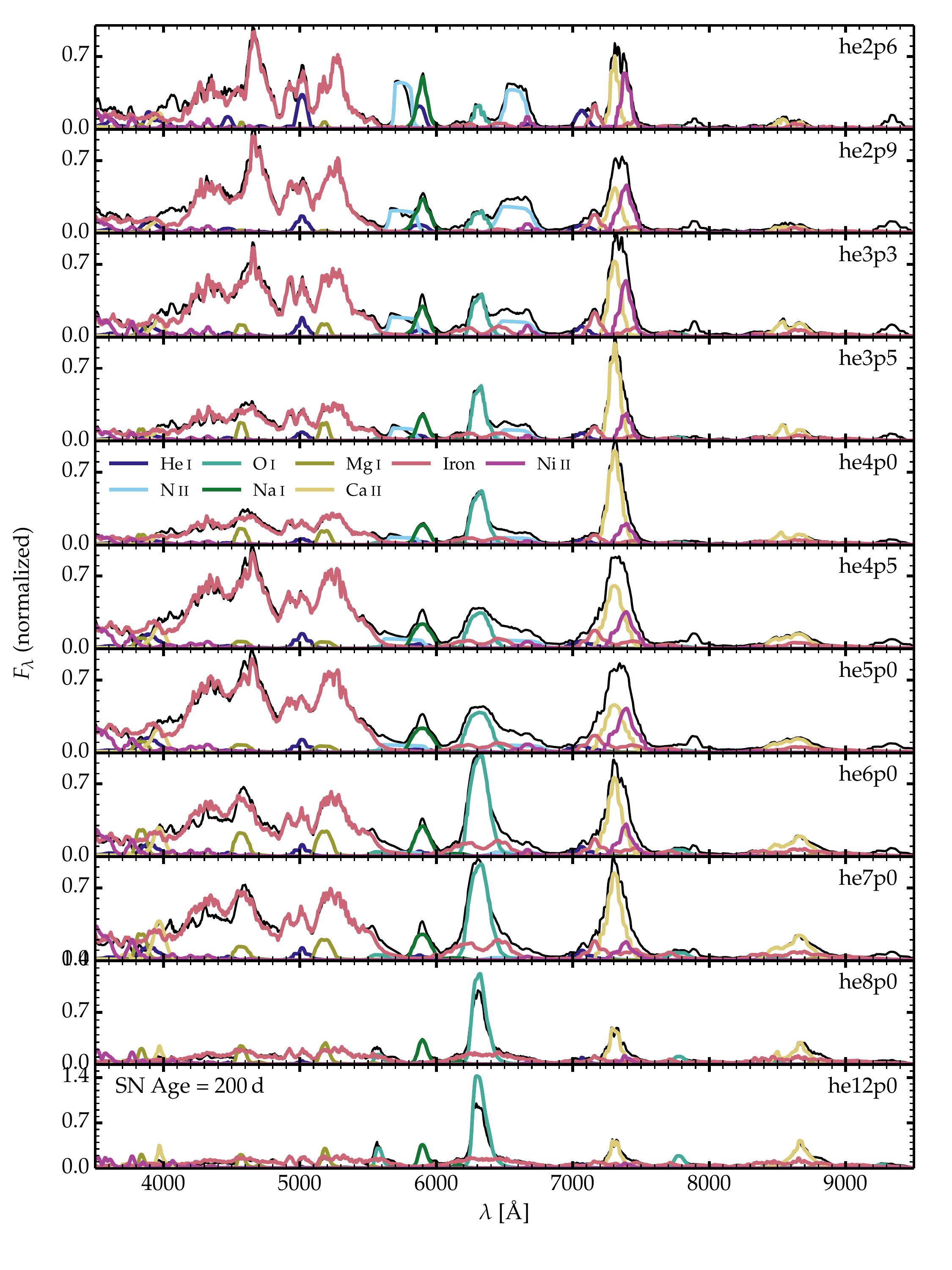}
\includegraphics[width=0.45\hsize]{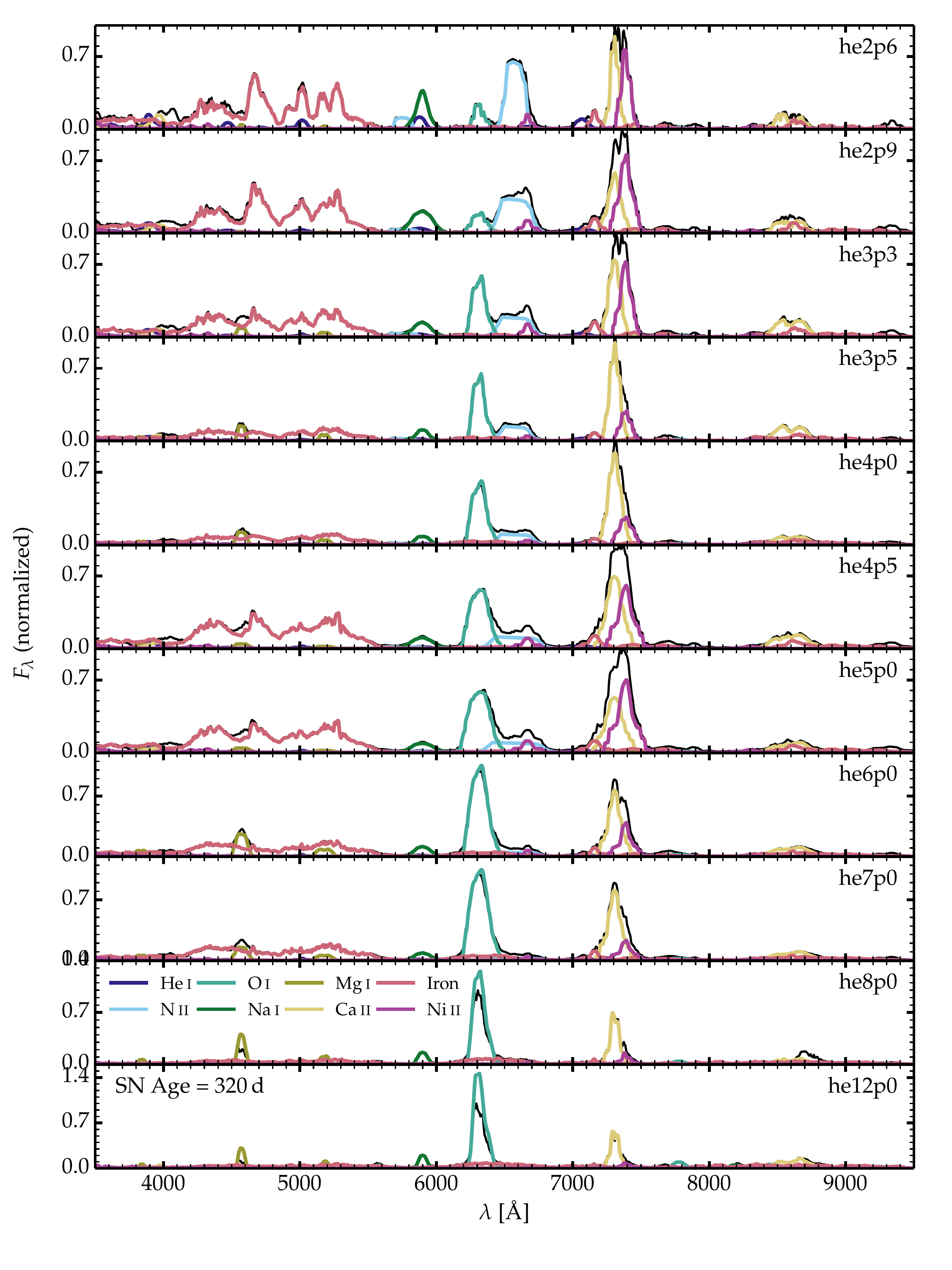}
\includegraphics[width=0.45\hsize]{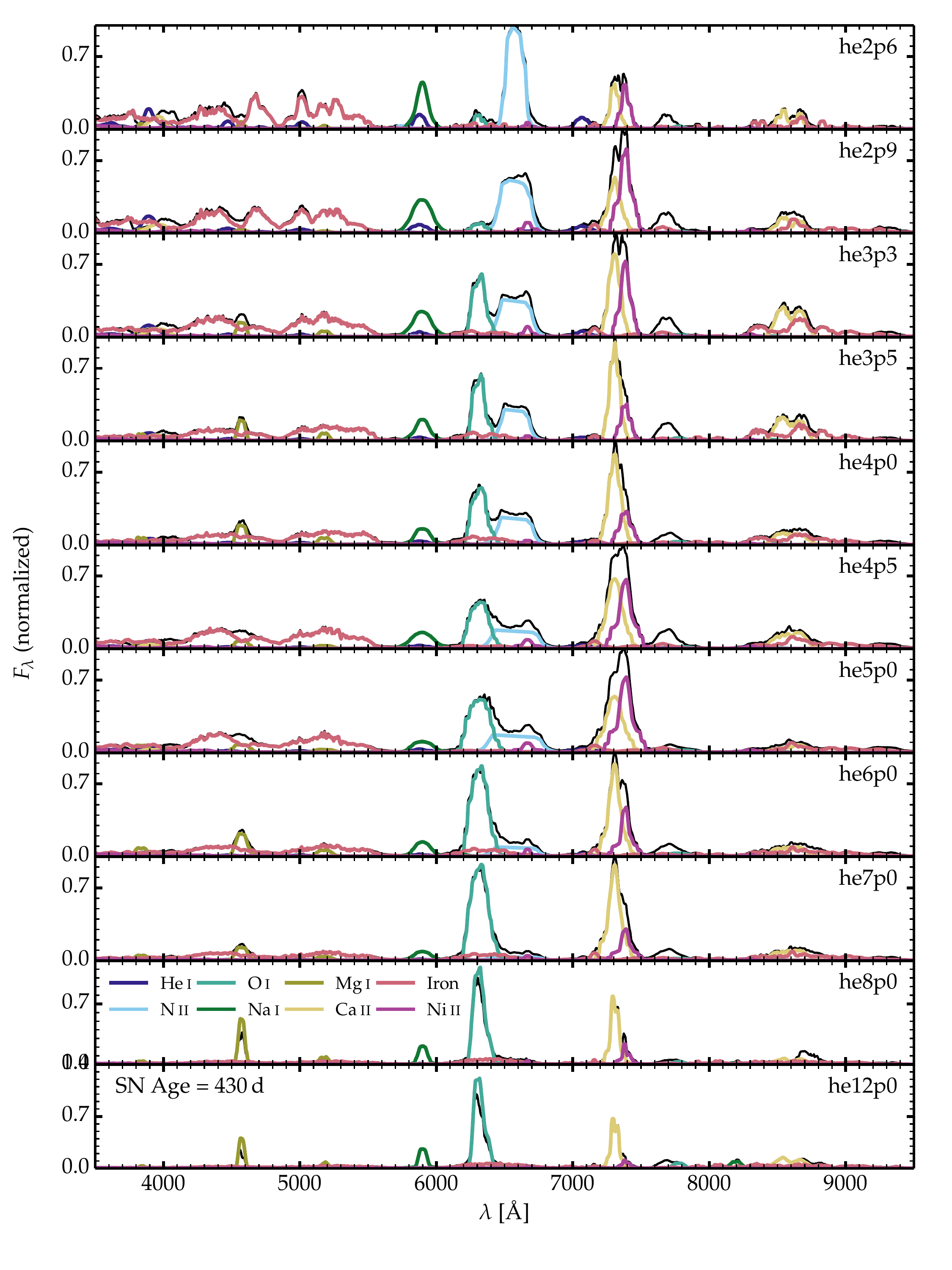}
\caption{Comparison of optical spectra for He-star explosion models he2p6 to he12p0 at a SN age of 100\,d (top left), 200\,d (top right), 320\,d (bottom left), and 430\,d (bottom right).
\label{fig_he_4epochs}
}
\end{figure*}

\begin{figure}
\centering
\includegraphics[width=\hsize]{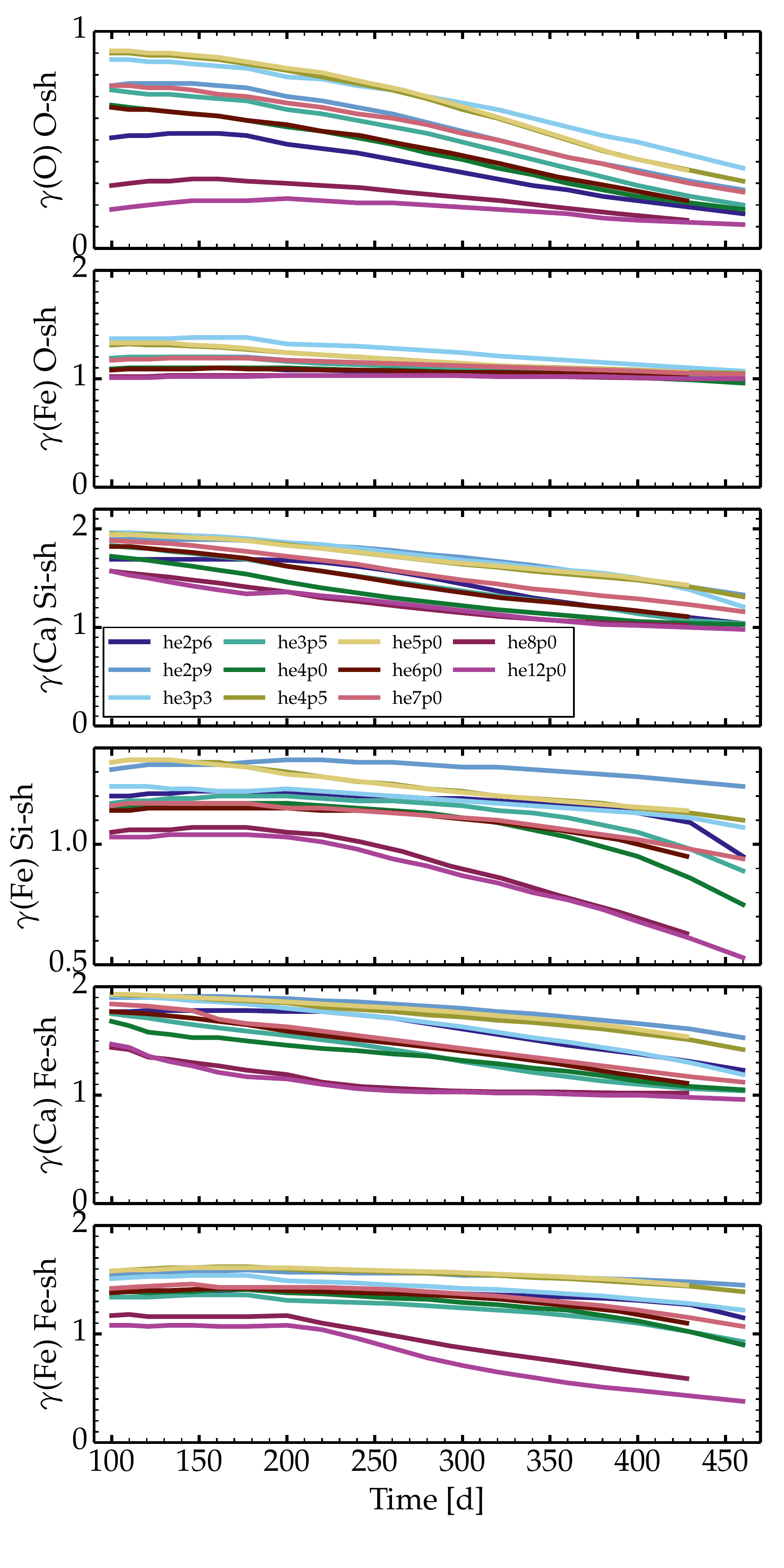}
\caption{Evolution of the ionization of important species in the O-rich, Si-rich, and Fe-rich shells in models he2p6 to he12p0. The ionization in the He-rich shell (not shown here) is roughly constant in time and such that He is mostly neutral, N is mostly N$^{+}$, and Fe is mostly Fe$^{2+}$.
\label{fig_ionization_evol}
}
\end{figure}

\begin{figure}
\centering
\includegraphics[width=\hsize]{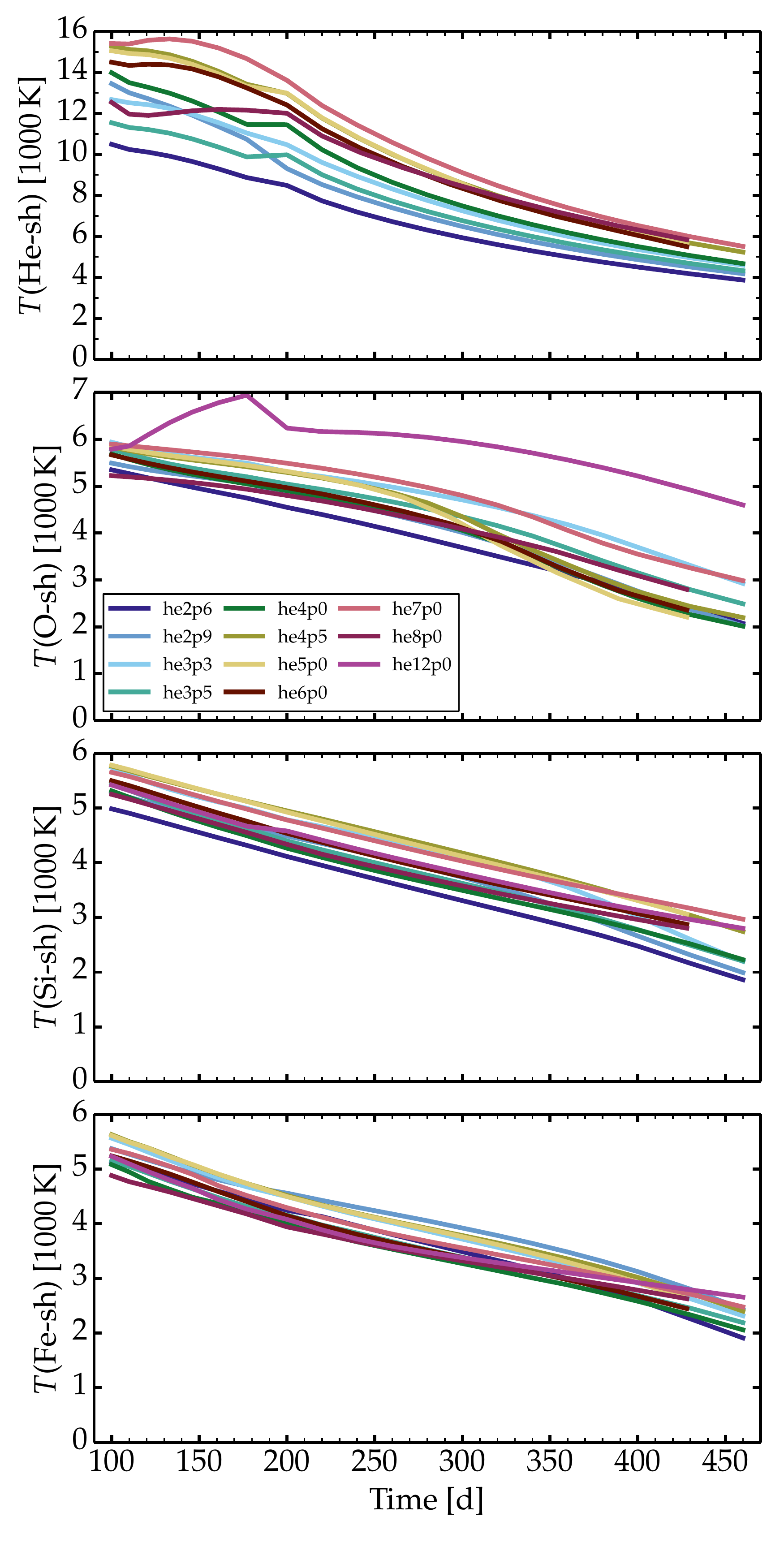}
\caption{Evolution of the temperature of important species in the He-rich, O-rich, Si-rich, and Fe-rich shells in models he2p6 to he12p0 (the he12p0 model does not have a He-rich shell and thus no temperature is output for that model in the top panel).
\label{fig_temp_evol}
}
\end{figure}

\end{appendix}

\end{document}

%% file: model_composition_table.tex
\begin{table*}
    \caption{Ejecta properties for our set of explosion models. The upper part describes the models already used in D21 while the lower part describes new models arising from progenitors evolved without wind mass loss.
\label{tab_prog}
    }
    \begin{center}
      \begin{footnotesize}
  \begin{tabular}{l@{\hspace{2mm}}|c@{\hspace{2mm}}c@{\hspace{2mm}}c@{\hspace{2mm}}
      c@{\hspace{2mm}}c@{\hspace{2mm}}c@{\hspace{2mm}}c@{\hspace{2mm}}c@{\hspace{2mm}}
      c@{\hspace{2mm}}c@{\hspace{2mm}}c@{\hspace{2mm}}c@{\hspace{2mm}}
      c@{\hspace{2mm}}c@{\hspace{2mm}}
    }
    \hline
Model  &  $M_{\rm ZAMS}$  & $M_{\rm preSN}$  & $M_{\rm ej}$  &     $E_{\rm kin}$   & $V_m$   &  \iso{4}He & \iso{12}C   & \iso{14}N  & \iso{16}O &  \iso{24}Mg &  \iso{28}Si &  \iso{40}Ca &  \iso{56}Ni$_{t=0}$ & \iso{58}Ni \\ 
       &     [\msun] & [\msun] & [\msun]   &        [foe]    & [\kms]    &    [\msun] & [\msun] & [\msun] & [\msun] & [\msun] & [\msun] & [\msun] & [\msun] & [\msun]  \\
\hline
   he2p6  & 13.85      & 2.15 &   0.79  &     0.13   &      4134   &      0.71    &      0.02  &   4.78(-3)   &   2.28(-2)   &   1.58(-3)   &   3.35(-3)   &   2.40(-4)   &   1.22(-2)  &  1.53(-3)   \\ 
   he2p9  & 14.82      & 2.37 &   0.93  &     0.37   &      6336   &      0.77    &      0.04  &   5.15(-3)   &   5.03(-2)   &   3.82(-3)   &   1.01(-2)   &   5.62(-4)   &   2.32(-2)  &  1.48(-3)   \\ 
   he3p3  & 16.07      & 2.67 &   1.20  &     0.55   &      6777   &      0.84    &      0.06  &   6.21(-3)   &   1.51(-1)   &   1.75(-2)   &   2.76(-2)   &   1.00(-3)   &   4.00(-2)  &  3.61(-3)   \\ 
   he3p5  & 16.67      & 2.81 &   1.27  &     0.41   &      5704   &      0.87    &      0.07  &   6.31(-3)   &   1.72(-1)   &   1.60(-2)   &   2.13(-2)   &   7.34(-4)   &   2.92(-2)  &  2.43(-3)   \\ 
   he4p0  & 18.11      & 3.16 &   1.62  &     0.63   &      6272   &      0.92    &      0.10  &   6.46(-3)   &   3.10(-1)   &   2.98(-2)   &   4.70(-2)   &   1.35(-3)   &   4.45(-2)  &  3.47(-3)   \\ 
   he4p5  & 19.50      & 3.49 &   1.89  &     1.17   &      7884   &      0.95    &      0.13  &   6.52(-3)   &   4.19(-1)   &   3.73(-2)   &   6.14(-2)   &   2.40(-3)   &   8.59(-2)  &  6.09(-3)   \\ 
   he5p0  & 20.82      & 3.81 &   2.21  &     1.51   &      8286   &      0.97    &      0.15  &   6.60(-3)   &   5.92(-1)   &   5.20(-2)   &   5.55(-2)   &   2.26(-3)   &   9.77(-2)  &  7.56(-3)   \\ 
   he6p0  & 23.33      & 4.44 &   2.82  &     1.10   &      6269   &      0.95    &      0.25  &   6.20(-3)   &   9.74(-1)   &   1.01(-1)   &   5.88(-2)   &   2.12(-3)   &   7.04(-2)  &  5.00(-3)   \\ 
   he7p0  & 25.68      & 5.04 &   3.33  &     1.38   &      6456   &      0.90    &      0.39  &   5.42(-3)   &   1.29(0)    &   1.07(-1)   &   9.47(-2)   &   3.42(-3)   &   1.02(-1)  &  4.19(-3)   \\ 
   he8p0  & 27.91      & 5.63 &   3.95  &     0.71   &      4251   &      0.84    &      0.49  &   5.17(-3)   &   1.71(0)    &   1.10(-1)   &   4.89(-2)   &   2.00(-3)   &   5.46(-2)  &  3.46(-3)   \\ 
   he12p0 & 35.74      & 7.24 &   5.32  &     0.81   &      3911   &      0.23    &      1.00  &   1.42(-4)   &   3.03(0)    &   8.73(-2)   &   7.41(-2)   &   3.42(-3)   &   7.90(-2)  &  2.47(-3)   \\ 
\hline
 he2p5MdZ  & 13.51 & 2.5  &   1.10  &     0.43  &     6237  &     0.87  &      0.05 &  4.08(-3) &  9.20(-2) &  3.41(-3) &  1.33(-2) &   8.39(-4) &  3.48(-2) &  2.58(-3)    \\ 
 he3p0MdZ  & 15.14 & 3.0  &   1.45  &     0.72  &     7077  &     0.98  &      0.08 &  7.12(-3) &  2.07(-1) &  1.29(-2) &  3.82(-2) &   2.25(-3) &  5.64(-2) &  2.51(-3)    \\ 
 he3p5MdZ  & 16.67 & 3.5  &   1.93  &     1.21  &     7941  &     1.06  &      0.10 &  7.34(-3) &  4.26(-1) &  4.25(-2) &  7.14(-2) &   3.57(-3) &  1.02(-1) &  5.02(-3)    \\ 
 he4p0MdZ  & 18.11 & 4.0  &   2.34  &     1.20  &     7168  &     1.11  &      0.13 &  7.45(-3) &  6.71(-1) &  5.88(-2) &  8.93(-2) &   4.53(-3) &  1.07(-1) &  4.75(-3)    \\ 
 he4p5MdZ  & 19.50 &  4.5  &   2.78  &     1.20  &     6573  &     1.15  &      0.16 &  6.60(-3) &  9.62(-1) &  7.91(-2) &  9.27(-2) &   4.33(-3) &  1.01(-1) &  4.44(-3)    \\ 
 \hline
  \end{tabular}
  \end{footnotesize}
\end{center}
    {\bf Notes:} The table columns correspond to the ZAMS mass, the preSN mass, the ejecta mass, the ejecta kinetic energy (1 foe $\equiv 10^{51}$\,erg), the mean
    expansion rate $V_{\rm m} \equiv \sqrt{2E_{\rm kin}/M_{\rm ej}}$, the cumulative yields of \iso{4}He, \iso{12}C, \iso{14}N, \iso{16}O, \iso{24}Mg,
    \iso{28}Si, \iso{40}Ca, \nifs\ prior to decay, and \iso{58}Ni. All \nifs\ masses used here were taken from the \kepler\ approximation
    to the \photb\ neutrino simulation (see discussion in D21).
\end{table*}